\documentclass[aps,floats,amssymb,twocolumn,prb, superscriptaddress]{revtex4-2}

\usepackage{calc}
\usepackage{psfrag}
\usepackage{graphicx}
\usepackage{color}
\usepackage{bm}
\usepackage{float}
\usepackage{lipsum}
\usepackage{amsmath}

\def\ds{\stackrel{\star}{,}}
\def\nn{\nonumber}
\def\bs{\boldsymbol}

\begin{document}

\title{Microscopic derivation of Dirac composite fermion theory: \\
Aspects of noncommutativity and pairing instabilities}

\author{Dragoljub Go\v{c}anin}
\affiliation{Faculty of Physics, University of Belgrade, Studentski Trg 12-16, 11000 Belgrade, Serbia}
\author{Sonja Predin}
\affiliation{Institute of Information Systems, Alfons-Goppel-Platz 1, 95030 Hof, Germany}
\author{Marija Dimitrijevi\'{c} \'{C}iri\'{c}}
\affiliation{Faculty of Physics, University of Belgrade, Studentski Trg 12-16, 11000 Belgrade, Serbia}
\author{Voja Radovanovi\'{c}}
\affiliation{Faculty of Physics, University of Belgrade, Studentski Trg 12-16, 11000 Belgrade, Serbia}
\author{Milica Milovanovi\'{c}}
\email{Corresponding author: milica.milovanovic@ipb.ac.rs}
\affiliation{Scientific Computing Laboratory, Center for the Study of Complex Systems,Institute of Physics Belgrade, University of Belgrade, Pregrevica 118, 11080 Belgrade, Serbia}


\begin{abstract}
Building on previous work [N. Read, Phys. Rev. B {\bf 58}, Z. Dong and T. Senthil, 16262 (1998); Phys. Rev. B {\bf 102}, 205126 (2020)] on the system of bosons at filling factor $\nu = 1$, we derive the Dirac
composite fermion theory for a half-filled Landau level from first principles and applying the Hartree-Fock approach in a preferred representation. On the basis of the microscopic formulation, in the long-wavelength limit, we propose a noncommutative field-theoretical description, which in a commutative limit reproduces the Son's theory, with additional terms that may be expected on physical grounds. The microscopic representation of the problem is also used to discuss
pairing instabilities of composite fermions. We find that a presence of a particle-hole symmetry breaking leads to a weak (BCS) coupling $p$-wave pairing in the lowest Landau level, and strong coupling $p$-wave
pairing in the second Landau level that occurs in a band with nearly flat dispersion, a third power function of momentum.
\end{abstract}

\maketitle

\section{Introduction}

The fractional quantum Hall effect (FQHE) can be explained by focusing on a strongly correlated problem of particles (electrons) in two dimensions when there is a commensuration between the number of particles and the number of flux quanta of the applied, orthogonal to the two-dimensional plane, magnetic field. Many phenomenological questions can be
answered by assuming that the most important physics takes place in a fixed Landau level (LL) with the precise
commensuration of the number of particles and the number of orbitals in a fixed LL (and other LLs are inert). That is why a mathematical, idealized problem of an isolated LL is so useful and relevant for the
understanding of the FQHE.

Some of the most interesting experimental phenomena occur at filling factors (ratio of the number of
electrons and the number of flux quanta) $\nu = \tfrac{1}{2}$ and $\nu =\tfrac{5}{2}$, even-denominator fractions. The gapless
system at $\nu =\tfrac{1}{2}$ is believed to be in a Fermi liquid state of underlying quasiparticles - composite fermions (CFs), as proposed and described in \cite{hlr} early on, while it is believed that incompressible (gapped) FQHE at $\nu = \tfrac{5}{2}$ can be
associated with some kind of $p$-wave pairing of CFs in the second LL (sLL), as
proposed in \cite{mr}. To understand more closely these systems, one may start by focusing on
an isolated half-filled LL, the lowest LL (LLL) at $\nu =\tfrac{1}{2}$ in the case of the CF liquid (CFL), and sLL at $\nu = \tfrac{5}{2}$ in the case of the gapped system.

One of the most interesting theoretical developments associated with the physics in an isolated, half-filled LL is the proposal in \cite{son} for the description of the CFL state that is based on an assumption that the
underlying quasiparticles: CFs can be effectively described as Dirac CFs, using
an effective Dirac theory in two dimensions. This can be of general interest: a system of interacting
fermions on a noncommutative (NC) space of an isolated LL, in which they fill half of the allowed, countable states, can be
described by an effective Dirac theory.

The proposed Dirac CF theory is a phenomenological theory, based on the assumption that an effective theory
of an isolated, half-filled LL must be manifestly invariant under the particle- (electron-) hole transformation. Certainly, there is a need for a microscopic derivation of the Dirac CF theory, which can serve as a base for further understanding of this strongly correlated system. In this paper we develop a
microscopic support for the Dirac CF theory, and provide a framework for a more detailed investigations. 

To describe the physics of an isolated half-filled LL, we 
generalize the approach to bosons in an 
isolated LL at filling factor one, of Pasquier and Haldane \cite{ph}, and later 
developed by Read \cite{read}, and more recently by Dong and 
Senthil \cite{dose}. This approach introduces additional, vortex degrees of freedom to efficiently capture Laughlin-Jastrow correlations. In the case of bosons, the vortex (holelike, 
unphysical) degrees of freedom are fermionic, and combine with 
elementary bosons to make quasiparticles of the problem - CFs. Due 
to their fermionic nature, the vortex degrees of freedom are 
uniformly distributed in the LL and make a uniform background. (This 
feature also guarantees a necessary independence of physics 
under transformations in the unphysical sector.) We generalize this 
description to the case of the half-filled LL of electrons, by 
assuming a uniform distribution of two kinds of unphysical degrees of 
freedom: holelike and electronlike. There are holelike vortices as 
many as particlelike (electronlike) vortices, and they behave as hard-core bosons 
among themselves, making a uniform background. To ensure the uniform 
background we need to introduce constraints in the description. 
Because of two kinds of unphysical degrees of freedom, the holelike 
vortices combine with electrons to make CFs, and the particlelike 
vortices combine with holes to make composite holes (CHs). We can 
choose either electrons or holes as physical degrees of freedom of the 
half-filled LL, but if we want to capture particle-hole (PH) symmetry (i.e., the symmetry under exchange of particles and holes), we should treat them on an equal footing. Thus, we need to include additional constraints that will preclude the simultaneous presence of a hole and an electron in an orbital of the fixed LL, and therefore describe them as dependent (not two independent) degrees of freedom. The requirement of the PH symmetry also justifies our assumption on the manner in which unphysical (vortex-like) degrees of freedom enter the description. In this way, on the basis of two sets of constraints, we are able to formulate the problem of the half-filled LL, that is explicitly invariant under exchange of particles and holes. Furthermore, within this framework, which explicitly includes the additional - vortex degrees of freedom, we are able to consider the ``preferred (form of) Hamiltonian" in the language of CFs and CHs, natural quasiparticles that, on the level of Hartree-Fock treatment, can effectively capture the physics of the system. Thus, a two-component fermion description, with the CF and CH fields, necessarily and naturally appears as a consequence of the demand for the PH symmetry, and we show that the description is of the Dirac type (when the constraints are taken into account). This provides a microscopic derivation of the Dirac description and explains the Dirac nature of the fermionic quasiparticle excitations of the low-energy physics. 

On the basis of this microscopic formulation, in the long-wavelength limit, we propose a noncommutative field-theoretical description, which in a commutative limit reproduces the Son's theory, with additional terms that may be expected on physical grounds. We also discuss pairing instabilities within the developed microscopic framework, and provide a physical understanding of the
$p$-wave pairing instability in the LLL, and in the sLL. The pairing in the LLL is of the BCS, weak-coupling
kind, and this may explain the scarcity of the pairing phenomena in the LLL. On the other hand, the pairing
in the sLL is of the strong coupling (weak pairing) kind as proposed and discussed in \cite{rg},
though we find that the Dirac CF band dispersion, $\epsilon(k)$, is flatter: it obeys a third-power law, i.e.,
$\epsilon(k)\sim k^3$.

The section that follows is a review of the bosonic problem at $\nu = 1$, in which we
also introduce a point of view of the formalism developed in \cite{read, ph},
that will be useful for the half-filled problem of electrons. Sections III and IV consider a (simpler) system, closely related to the one of the half-filled LL, a special-bilayer system with two kinds of
particles, parallel to the existence of electrons and holes in the half-filled LL. A transformation into
holes of just one kind of particles in the special-bilayer system enables a formulation of the half-filled
LL problem in Sec. V, with all necessary constraints. Following the usual approach \cite{ms} to a
formulation with constraints [that enables a Hartree-Fock (HF) treatment] we discuss a ``preferred" form of the
Hamiltonian in Sec. VI, and in Sec. VII a Dirac form of the Hamiltonian in the HF approximation. In Section VIII we describe how in the long-wavelength limit of the microscopic formulation we can reach a field-theoretical description with gauge fields next to the Dirac composite fermions. In Secs. IX and X we discuss the
description of possible pairing instabilities. The structure of the proposed NC field theory is described in Appendix A, while Appendix B concerns some specific aspects of the relevant covariant derivatives. Conclusions are summarized in Sec. XI.


\section{Review of the $\nu = 1$ boson system and introductory remarks}

A CF is a composite object, a bound state of an underlying elementary particle with a whole number of vortices; a vortex represents an excitation of the FQHE system due to an insertion of one flux quantum
that induces a depletion of charge. At filling factors $\nu =1/q$, where $q$ is an integer,
a composite fermion is a neutral object; a composite of an electron (fermion) and a hole (more precisely
a depletion of charge) associated with $q$ flux quanta, when $q$ is even, and a composite of a boson and
a hole associated with $q$ flux quanta, when $q$ is odd. To simplify the terminology, we will always call the excitation with $q$ flux quanta a vortex. These introductory remarks serve just to remind the
reader of the physical picture of the CF, and for an elaborate introduction to the CF formulation the
reader may consult \cite{read}. We conclude that it may be expected and natural that an operator describing annihilation or creation of CF will carry two indices, one for the state of the
elementary particle and the other for the state of the hole in an orthonormal basis. In the following
we will introduce the two-index formalism that
was firstly proposed in
\cite{ph} and further elaborated in \cite{read} and \cite{dose}.

We start from an enlarged space with (composite) fermion $c_{mn}^{\dagger}$ with two indices, each
corresponding to an orbital in the LLL (fixed LL): $m,n = 1,\dots, N_{\phi}\equiv N$ such that
\begin{equation}
\{c_{nm}, c_{m'n'}^{\dagger}\}= \delta_{n,n'} \delta_{m,m'}.
\end{equation}
Each $c_{nm}$ fermion represents a composite object. We define physical subspace of bosonic states in the
LLL by
\begin{equation}
\vert n_{1},\dots,n_{N}\rangle = \sum\limits_{m_{1},\dots,m_{N}}^{N_{\phi}} \varepsilon^{m_{1} \cdots m_{N}}
c_{m_{1}n_{1}}^{\dagger} \cdots c_{m_{N} n_{N}}^\dagger \vert 0\rangle
\label{bsts}
\end{equation}
where $\varepsilon^{m_{1} \cdots m_{N}}$ is the Levi-Civita symbol. In this way, bosonic physical states have
a property, defined by
\begin{equation}
\rho_{m m'}^{R} = \sum_{n} c_{m n}^{\dagger} c_{n m'},
\label{rgen}
\end{equation}
that
\begin{equation}
\rho_{m m}^{R}\vert n_{1},\dots,n_{N}\rangle  = 1 \cdot \vert n_{1},\dots,n_{N} \rangle,
\label{reqb2}
\end{equation}
expressing a single occupancy of each unphysical orbital $m$. The physical states are defined by this property
so that unphysical orbitals make an uniformly occupied background. They furnish a spin-singlet representation of $SU(N)$ group,
\begin{equation}
m \neq m' , \; \; \; \rho_{m m'}^{R}\vert n_{1},\dots,n_{N}\rangle = 0,
\label{reqb1}
\end{equation}
which is in agreement with the requirement that the physics should not depend on the choice of basis in
the unphysical sector $R$.

But we may reformulate this requirement by demanding that (1) $\rho_{mm}^R = 1$  and that (2) ``unphysical" particles (objects), which are uniformly
distributed in the LLL, are fermions. This will automatically lead to (\ref{reqb1}).  In other words, introducing a certain  type of statistics in the unphysical sector is
a way of specifying physical states together with the demand  for $\rho_{mm}^R = 1$.  In the following we would like to further this view and introduce an approach, which 
will make a basis for our description of the $\nu = \tfrac{1}{2}$ problem, and contrast it with respect to the previous construction of the bosonic $\nu = 1$ state. 

First, we will recapitulate the Pasquier-Haldane construction, and the way vortex statistics enters the description. The  Pasquier-Haldane enlarged space consists of states,
\begin{equation}
c_{m_{1}n_{1}}^{\dagger} \cdots c_{m_{N} n_{N}}^\dagger \vert 0\rangle,
\end{equation}
which describe the presence of $N$ composite objects. To each object we associate two single-particle LL states, the state $n_i $ of the elementary, physical particle - boson, and state $m_i $ of vortex. By exchanging states, 
$n_i  \leftrightarrow n_j $, or $m_i  \leftrightarrow m_j $ , we do not get the same state up to a sign, i.e., we cannot speak about definite statistics. But if we trace out one of the (two) degrees of freedom, like by tracing out in an anti-symmetric way vortices in (\ref{bsts}), we can speak about definite statistics; we have to do an ordinary (symmetric) trace in all $n$ indices to get a state with definite statistics in the unphysical sector - a single Slater (Vandermonde) determinant of fermionic vortices in this $\nu = 1$ case. 

This motivates our approach: we consider an enlarged space (a subspace of the Pasquier-Haldane space) in which all $m$'s (in the unphysical sector) are different (i.e., $m_i  \neq m_j $  for $ i \neq j ; i,j = 1, \ldots, N ). $ Physically, we may understand this as a
modeling of a uniformly distributed vortex background. The $SU(N)$ invariance in the $R$ sector is  present, and acts trivially, because each generator, (\ref{rgen}) with $m \neq m'$ will map out of this restricted space, and thus effectively  (considering the necessary projection) annihilate state in the restricted space. If we further require that the states of the restricted space have definite (fermionic) statistics of vortices we will have a unique realization of the $SU(N)$ symmetry, given by (\ref{bsts}).  Thus, we do not constrain the unphysical degrees of freedom by  a single symmetry requirement (as in the approach to the bosonic $\nu = 1$ in \cite{dose}, and which is possible, allowed in that system), but primarily by a physical requirement of the uniform vortex (unphysical degrees of freedom) density. Thus, we consider a description with an (additional but necessary) constraint in the unphysical sector, which is present automatically in the bosonic $\nu = 1$ case, because of the fermionic statistics of the unphysical degrees of freedom. The constraint of the uniform vortex density is a way to introduce bosonic vortices into a theory, and still maintain the $SU(N)$ invariance in the unphysical (vortex) sector, i.e., invariance of the physics under change of basis (transformations) in the unphysical sector. We will consider examples of this at the end of this section and in the following section when we discuss a special bilayer problem as a preparation for the half-filled LL problem. Furthermore, we will see that in the half-filled LL problem, unphysical degrees of freedom are associated with both $L$ and $R$ sectors, and thus we see that in that case also, the requirement of the uniform density of the unphysical degrees of freedom  is the most natural to constrain these degrees of freedom, and maintain the invariance under change of basis [not by a spin-singlet realization of $SU(N)$ symmetry in one ($R$) sector]. 
Therefore, we can specify physical states by demanding the uniform distribution of unphysical degrees of freedom in the states that enter description of the physical states.
These states form a  space that is  effectively invariant under  $SU(N)$ transformations. The demand for a definite statistics of unphysical degrees of freedom in these states leads to physical states.

Thus, in principle, we can discuss a possible physical sector (system) in an enlarged theory for which
$\rho_{mm}^R=1$, but with the unphysical particles being (``hard-core") bosons, and the physical sector being a $\nu=1$ fermionic system. The old requirement (\ref{reqb1}) would be fulfilled effectively (= under the necessary projection) in a restricted space for which $\rho_{mm}^R = 1$, i.e., space defined by {\em hard-core} vortex configurations (of $N$ of vortices). The physical space is defined by a hard-core configuration of vortices that correlate as bosons.
In this case 
\begin{equation}\label{eq7}
\vert n_{1},\dots,n_{N} \rangle_f = \sum\limits_{m_{1},\dots,m_{N}}^{N_{\phi}} s^{m_{1} \cdots m_{N}}
c_{m_{1}n_{1}}^{\dagger} \cdots c_{m_{N} n_{N}}^\dagger \vert 0\rangle,
\end{equation}
where
$s^{m_{1} \cdots m_{N}} = 1$ only if no index is equal to
any other index, but otherwise zero.
But as we know, this would not lead to a plausible Hartree-Fock description, i.e., a good representation in which the Hartree-Fock approach
to the system of composite fermions $c_{n m}$ would make a good starting point for more refined descriptions.

Therefore, in principle, we can consider both, bosonic
and fermionic realizations of the $SU(N)$ group (that
works in the unphysical sector), by considering the theory with CFs as building blocks of the effective description, at a particular fraction, and only one will realize as
a physical theory. [We should choose composite bosons
(CBs) in the system in which the CF realization fails.]
The single occupancy in the unphysical sector will lead
to definite statistics (fermionic or bosonic) in the unphysical sector, because the algebra of the density operators (that the theory is built on), in a LL basis, is
\begin{equation}\label{SU(N)_algebra}
[\rho^{R}_{mm'},\rho^{R}_{ll'}]=\rho^{R}_{ml'}\delta_{m',l}-\rho^{R}_{lm'}\delta_{m,l'},
\end{equation}
i.e., the algebra of $SU(N)$ generators that can be realized
either by fermions or bosons. The constant density and
demand for fermionic statistics will coincide with a simple
$SU(N)$ invariance: the group action will map the state
in (\ref{bsts}) into itself. The constant density requirement and
demand for bosonic statistics will coincide with a special
$SU(N)$ invariance (a maintenance of the symmetry under the projection to the constant density): the group
action will map the state in (\ref{eq7}) into the same state up
to a number (a coefficient because of the projection) and
only the symmetric group $S_{n}$, for $n=N$, a subgroup of
$SU(N)$, represented by signed permutation matrices, will
map the state into itself, up to a sign. Thus, by fixing the density,
we will have either fermionic or bosonic realization in the
unphysical sector, and in the following we will emphasize
which realization (statistics) we choose.

\section{Towards $\nu = \tfrac{1}{2}$ fermions - a special bilayer system}

As a preparation for a set-up of the $\nu = \tfrac{1}{2}$ (i.e. the half-filled LL) problem in the language of composite quasiparticles, we will discuss a set-up for a special quantum Hall bilayer system. The special bilayer is characterized by two layers of electrons, each at filling factor $\nu = \tfrac{1}{2}$ of a fixed LL, but what is special is that an electron in one layer cannot be in the same (LL) orbital with an electron from the other layer. Thus, we may speak about two kinds of composite objects, composite fermions $c$ and $d$, which are overall neutral objects consisting of elementary particle - electron of a given layer, and unphysical (vortex) object of opposite charge. Therefore, we consider two CFs, $ c^{\dagger}_{m n}$ and $ d^{\dagger}_{m n}$, for the two layers, and an enlarged space with states
\begin{equation}
c_{m_{1}n_{1}}^{\dagger} \cdots c_{m_{N/2} n_{N/2}}^\dagger d_{m'_{1}n'_{1}}^{\dagger} \cdots d_{m'_{N/2} n'_{N/2}}^\dagger\vert 0\rangle,
\end{equation}
i.e., always there are $N/2$ $c$-fermions and $N/2$ $d$-fermions, where $N$, as before, is the
number of available orbitals in the LL.

The density of the unphysical objects, vortices (of the opposite charge with respect to electrons), may be expressed in the analogous way as in the previous section:
\begin{equation}
\rho_{m m'}^{R(c)} = \sum_{n} c_{m n}^{\dagger} c_{n m'},
\end{equation}
and
\begin{equation}
\rho_{m m'}^{R(d)} = \sum_{n} d_{m n}^{\dagger} d_{n m'}.
\end{equation} 

Following the discussion in the previous section, we require
\begin{equation}
\rho_{m m}^{R(c)}+ \rho_{m m}^{R(d)} = 1,
\label{reqf1}
\end{equation}
i.e., we uniformly distribute particles in the unphysical sector. Furthermore, we choose them to be bosons
and mutual bosons. The requirement (\ref{reqf1}) may be associated with special (identical for both $c$-fermions and
$d$-fermions) transformations of the LL basis in the unphysical sector, which we denote by $SU^{R}_{c}(N)$ (where $c$ stands for charge), a transformation that is realized identically on both $c$-fermions and $d$-fermions by affecting
their unphysical index. These transformations would leave the physical states unchanged if we work in the restricted subspace with (unphysical) hard-core bosons
\begin{equation}
m \neq m' , \;  \left(\rho_{m m'}^{R(c)} + \rho_{m m'}^{R(d)}\right) \vert n_{1},\dots,n_{N/2} n'_{1},\dots,n'_{N/2}\rangle = 0,
\end{equation}
where the equality is the result of the projection to the restricted state,
and thus, by assuming (\ref{reqf1}) and bosonic correlations in the unphysical sector, we have that
\begin{eqnarray}
&& \vert n_{1},\dots,n_{N/2},n'_{1},\dots,n'_{N/2}\rangle  \nonumber \\
&& =\sum\limits_{m_{1},\dots,m_{N/2}, m'_{1},\dots,m'_{N/2}}^{N} s^{m_{1} \cdots m_{N/2} m'_{1},\dots,m'_{N/2}} \times \nonumber \\
&& c_{m_{1}n_{1}}^{\dagger} \cdots c_{m_{N/2} n_{N/2}}^\dagger d_{m'_{1}n'_{1}}^{\dagger} \cdots d_{m'_{N/2} n'_{N/2}}^\dagger \vert 0\rangle,
\label{enstates}
\end{eqnarray}
where $s^{m_{1} \cdots m_{N/2} m'_{1},\dots,m'_{N/2}}$ is non-zero, equal to one only if no index is equal to
any other index.

In (\ref{enstates}) we have not only required that the unphysical - bosonic degrees of freedom are
uniformly distributed (\ref{reqf1}), but that they correlate mutually in a symmetric way. Thus, we chose
a sector of definite statistics in the enlarged space.

If $N = 2$, we have the following candidates for physical states:
\begin{equation}
\vert n, n'\rangle = (c_{1n}^{\dagger} d_{2n'}^{\dagger} + c_{2n}^{\dagger} d_{1n'}^{\dagger})\vert 0\rangle,
\end{equation}
where $ n, n' = 1, 2$. Additionally, as a part of the definition, we require that in the special bilayer system, i.e.,
two half-filled LL system, LL orbitals in the physical states cannot be doubly occupied. Thus, what is needed is
to suppress the unwanted states (those with $(n, n') = (1, 1)$ or $(2, 2)$ in the $N=2$ example) of double occupancy (with
an eye on the half-filled problem). Therefore, we need also
\begin{equation}
\rho_{n n}^{L(c)}+ \rho_{n n}^{L(d)} = 1,
\label{reqf2}
\end{equation}
This leads to $\vert\text{phy}\rangle$, physical states for which
\begin{equation}
n \neq n' , \;\; \left(\rho_{n n'}^{L(c)}\pm \rho_{n n'}^{L(d)}\right)\vert\text{phy}\rangle = 0.
\label{physector}
\end{equation}
We may associate the plus combination with $SU^{L}_{c}(N)$, and the minus combination with
$SU^{L}_{s}(N)$ - ``spin" transformations - which are inverse in the $d$-sector with respect to the ones
in the $c$-sector. Together, (\ref{reqf2}) and (\ref{physector}) with the plus sign lead to conclusion that $\vert\text{phy}\rangle$
states are spin singlet(s) under  $SU^{L}_{c}(N)$. On the other hand, in the physical states, the generators of $SU^{L}_{s}(N)$
transformations,
\begin{equation}
\rho_{n n}^{L(c)} - \rho_{n n}^{L(d)},
\end{equation}
may have expectation values from the interval $[-1, 1]$. At the beginning, before the requirement (\ref{reqf2}), operator sets, $\{\rho_{n n'}^{L(c)}\}$ and $\{\rho_{n n'}^{L(d)}\}$, with the constraints $\sum \rho_{n n}^{L(c)} = \sum \rho_{n n}^{L(d)} = N/2$, furnished
two adjoint representations of $SU(N)$ group. However, with the hard-core constraint (\ref{reqf2}), we have only one
non-trivial representation of $SU(N)$ group, $SU^{L}_{s}(N)$. The physical states are invariant under global
$U_{s}(1)$ transformation because
\begin{equation}
\sum_{n} \rho_{n n}^{L(c)} = \sum_{n} \rho_{n n}^{L(d)} = N/2,
\label{reqf3}
\end{equation}
and so are the unphysical ($R$-sector) states,
\begin{equation}
\sum_{n} \rho_{n n}^{R(c)} = \sum_{n} \rho_{n n}^{R(d)} = N/2,
\label{reqf4}
\end{equation}
by definition.

This completes a constraint [(\ref{reqf1}), (\ref{reqf2}), (\ref{reqf3}), (\ref{reqf4})] plus statistics set-up for the description in an enlarged space of the problem that concerns a special bilayer at $\nu_{\rm{tot}} = 1$
with two kinds of electrons, i.e., composite $c$-fermions and $d$-fermions, which cannot occupy the same orbital in
the restricted space of a fixed LL.

\section{The special bilayer system and its transformation into the half-filled system}

We may consider the previous formulation of the special bilayer as a starting point for the formulation of the half-filled problem. To reach the half-filled problem, one kind of electrons (in one of the layers) should transform, i.e., become (elementary, physical) holes. If we consider a layer with composite $d$-fermion, under the special electrons into holes transformation, what was the density of ``physical"  electrons, $ \rho^{L(d)}$, should become the description of the density of unphysical vortices, quasielectrons, and be a part of the constraints and description in the unphysical sector, i.e., $\rho^{R(d)} \rightarrow \rho^{L(d)}$. And similarly, what was the density of unphysical vortices (with charge opposite to the one of electron) $ \rho^{R(d)}$, should become the description of the density of ``physical" holes, and be a part of constraints and description in the physical sector, i.e., $\rho^{L(d)} \rightarrow\rho^{R(d)}$. 

Thus, by making a PH transformation, i.e., charge conjugation in one of the layers, described by composite object $d$, it is appropriate to call this object {\em composite hole}, because now the physical degree of freedom is a hole. 

We may make this discussion more concrete by considering the special bilayer description in the inverse space and fixing the notation that will be in place also for the half-filled case. Also, we will discuss possible particle-hole transformations in the special bilayer case and derive again the necessary transformations that transform the special bilayer problem into the one of the half-filled LL. 
 
 
We consider a representation of $c$ and $d$ composite fermions of the special bilayer in the inverse (momentum) space.
Following the previous studies on the $\nu = 1$ bosonic problem, we introduce the following decompositions:
\begin{equation}
c_{n m} = \int\frac{d\bs{k}}{(2 \pi)^{\frac{3}{2}}} \langle n\vert\tau_{\bs{k}}\vert m\rangle c_{\bs{k}},
\end{equation}
and
\begin{equation}
d_{n m} = \int \frac{d\bs{k}}{(2 \pi)^{\frac{3}{2}}} \langle n\vert\tau_{\bs{k}}\vert m\rangle d_{\bs{k}},
\end{equation}
with $\tau_{\bs{k}} = \exp\left(i \bs{k} \cdot\bs{R}\right)$, where $\bs{R}$ is a guiding-center coordinate of a single
particle in the external magnetic field, 
\begin{equation}
[R_x , R_y ] = - i ,
\end{equation}
we took $l_B$ (magnetic length) $=1$,
and $\{\vert n\rangle\}$ are single-particle states (orbitals) in a fixed LL.

With these decompositions we find that
\begin{equation}
\rho_{n n'}^{L(c)} = \sum_m c_{m n}^{\dagger}c_{n' m} = \int \frac{d\bs{q}}{2\pi} \langle n'\vert \tau_{\bs{q}}\vert n\rangle
\rho_{\bs{q}}^{L(c)},
\label{den}
\end{equation}
where
\begin{equation}
\rho_{\bs{q}}^{L(c)} = \int \frac{d\bs{k}}{(2\pi)^2} c_{\bs{k} - \bs{q}}^\dagger c_{\bs{k}} \exp\left(i \frac{\bs{k} \times \bs{q}}{2}\right), \label{bden}
\end{equation}
and similarly for $\rho_{n n'}^{L(d)}$,
\begin{equation}
\rho_{n n'}^{L(d)} = \sum_m d_{m n}^{\dagger}d_{n' m} = \int \frac{d\bs{q}}{2\pi} \langle n'\vert \tau_{\bs{q}}\vert n\rangle
\rho_{\bs{q}}^{L(d)},
\label{dend}
\end{equation}
where
\begin{equation}
\rho_{\bs{q}}^{L(d)} = \int \frac{d\bs{k}}{(2\pi)^2} d_{\bs{k} - \bs{q}}^\dagger d_{\bs{k}} \exp\left(i \frac{\bs{k} \times \bs{q}}{2}\right). \label{bdend}
\end{equation}
Note the inverse order of indices, $n$ and $n'$, on the left- and
right-hand sides of (\ref{den}) and (\ref{dend}). Similarly, 
\begin{equation}
\rho_{m m'}^{R(c)} = \sum_n c_{m n}^{\dagger}c_{n m'} = \int \frac{d\bs{q}}{2\pi} \langle m\vert \tau_{\bs{q}}\vert m'\rangle
\rho_{\bs{q}}^{R(c)},
\label{denr}
\end{equation}
where
\begin{equation}
\rho_{\bs{q}}^{R(c)} = \int \frac{d\bs{k}}{(2\pi)^2} c_{\bs{k} - \bs{q}}^\dagger c_{\bs{k}} \exp\left(- i \frac{\bs{k} \times \bs{q}}{2}\right), \label{bdenr}
\end{equation}
and analogously for $\rho_{n n'}^{R(d)}$,
\begin{equation}
\rho_{m m'}^{R(d)} = \sum_n d_{m n}^{\dagger}d_{n m'} = \int \frac{d\bs{q}}{2\pi} \langle m\vert \tau_{\bs{q}}\vert m'\rangle
\rho_{\bs{q}}^{R(d)},
\label{denrd}
\end{equation}
where
\begin{equation}
\rho_{\bs{q}}^{R(d)} = \int \frac{d\bs{k}}{(2\pi)^2} d_{\bs{k} - \bs{q}}^\dagger d_{\bs{k}} \exp\left(- i \frac{\bs{k} \times \bs{q}}{2}\right), \label{bdenrd}
\end{equation}
Note the positions of indices $m$ and $m'$ on the left- and right-hand sides of (\ref{denr}) and (\ref{denrd}).

We have
\begin{equation}
[\rho_{\bs{q}}^{L}, \rho_{\bs{q}'}^{L}] = 2 i \sin\left(\frac{\bs{q}\times\bs{q}'}{2}\right)
 \rho_{\bs{q} + \bs{q}'}^{L},
\end{equation}
and
\begin{equation}
[\rho_{\bs{q}}^{R}, \rho_{\bs{q}'}^{R}] = - 2 i \sin\left(\frac{\bs{q}\times\bs{q}'}{2}\right)
 \rho_{\bs{q} + \bs{q}'}^{R},
\end{equation}
i.e., Girvin-MacDonald-Plazmann (GMP) algebra for two kinds of particles - particles with opposite electric charge.

We may introduce a particle-hole transformation in the $d$-sector by taking
\begin{equation}
d_{\bs{k}} \rightarrow d_{-\bs{k}}^\dagger \hspace{0.2cm} (d_{\bs{k}}^\dagger \rightarrow d_{-\bs{k}}).
\end{equation}
This implies
\begin{equation}
d_{m n} \rightarrow d_{m n}^\dagger \hspace{0.2cm} (d_{m n}^\dagger \rightarrow d_{m n}),
\end{equation}
and also for
\begin{equation}
\rho_{n n'}^{L(d)} = \sum_m d_{m n}^\dagger d_{n' m}, \;\; n \neq n'
\end{equation}
we have
\begin{equation}
\rho_{n n'}^{L(d)} \rightarrow \sum_m d_{m n} d_{n' m}^\dagger = - \sum_m d_{n' m}^\dagger d_{m n} = - \rho_{n' n}^{R(d)}.
\label{pht}
\end{equation}
In the inverse space this implies
\begin{equation}
\rho^{L(d)}({\bs{q}}) \rightarrow - \rho^{R(d)}(\bs{q}),
\end{equation}
which is consistent with our expectation of what a particle-hole transformation should imply on the physical density in the $d$-sector; it should induce a density of particles of opposite charge in the magnetic field
($L \rightarrow R$) (and a minus sign that is always accompanied with such a transformation). We can reach
the same result by considering $\rho^{L(d)}(\bs{q})$ for ${\bs{q}} \neq 0$,
\begin{equation}
\rho^{L(d)}({\bs{q}})= \int \frac{d{\bs{k}}}{(2 \pi)^2} d_{{\bs{k}} - {\bs{q}}}^\dagger d_{{\bs{q}}} \exp\left(i
\frac{{\bs{k}}\times {\bs{q}}}{2}\right),
\end{equation}
and applying the transformation $ d_{\bs{k}} \rightarrow d_{-\bs{k}}^\dagger$ $(d_{\bs{k}}^\dagger \rightarrow d_{-\bs{k}})$. 

Therefore, this transformation may be identified to be the one that corresponds
(in the enlarged space) to the particle-hole transformation on the elementary (fundamental) degrees of
freedom, $e_{d}$, the second kind of electrons in the special bilayer: $e_{d} \rightarrow h_{d}^\dagger$,
$e_{d}^\dagger \rightarrow h_{d}$ (where $e_d$, $e_d^\dagger$, $h_d$, $h_d^\dagger$ are annihilation and
creation operators).

Above we introduced the effect of the particle-hole transformation (on electrons $e_d$) in the $d$-sector
on composite fermion operators, while putting aside the question of the diagonal terms, $\rho_{n n}^{L(d)}$, and the necessary existence of a constant term, equal to $N$, due to the anti-commutation relation of
$d_{m n}$'s. This would imply an additional delta-function contribution ($\sim \delta^{2}(\bs{q})$) in the
inverse space for the particle-hole transformation of $\rho^{L(d)}(\bs{q})$.

To comply with the restrictions of physical spaces in the enlarged spaces of composite fermion operators,
$d_{n m}$ and $c_{n m}$, of the half-filled and special bilayer problem, we expect
\begin{equation}
\rho_{n n}^{L(d)} \rightarrow 1 - \rho_{n n}^{R(d)},
\end{equation}
because the summation on $n$ on both sides, and the restrictions
\begin{equation}
\sum_n \rho_{n n}^{L(d)} = \sum_n \rho_{n n}^{R(d)} = \frac{N}{2},
\end{equation}
would be consistent with the particle-hole (single-layer) symmetry of the physical system and
restrictions on the special bilayer system.

Thus, in order to project the transformation $d_{m n} \rightarrow d_{m n}^\dagger $ on the physical spaces
of half-filled and special bilayer problems we demand
 \begin{equation}
 \rho_{n n}^{L(d)} \rightarrow 1 - \rho_{n n}^{R(d)},
\end{equation}
which requires an additional subtraction of a constant term $(N - 1)$ after the $d_{m n} \rightarrow d_{m n}^\dagger$ $(d_{m n}^\dagger \rightarrow d_{m n})$ transformation, in order to project out the unphysical
degrees of freedom. In the inverse space this affects the delta function contribution; thus, for ${\bs{q}}\neq 0$ we still have $\rho^{L(d)}({\bs{q}}) \rightarrow - \rho^{R(d)}(\bs{q})$.

In the following we would like to examine how this particle-hole transformation affects the constraints
imposed on the bilayer system in order to see how they look like in the (enlarged) space of the special
bilayer system. The two ``hard-core" constraints, $\rho_{n n}^{R(c)} + \rho_{n n}^{R(d)} = 1$ in (\ref{reqf1}) and  $\rho_{n n}^{L(c)} + \rho_{n n}^{L(d)} = 1$  in (\ref{reqf2}) become
\begin{equation}
 \rho_{n n}^{R(c)} = \rho_{n n}^{L(d)},
\end{equation}
and
\begin{equation}
 \rho_{n n}^{L(c)} = \rho_{n n}^{R(d)},
\end{equation}
respectively, which is consistent with the view that now the $d$-sector is described by hole degrees of freedom. On the other hand, the operator $\rho_{n n}^{L(c)} - \rho_{n n}^{L(d)} $ transforms into $\rho_{n n}^{L(c)} + \rho_{n n}^{R(d)} - 1$, i.e., $\rho_{n n}^{L(c)} + \rho_{n n}^{R(d)} $ acquires expectation values in the physical states ranging from $0$ to $2$. Thus, we introduced the hole view in the $d$-sector and this increased the allowed occupancy of $c$-particles and $d$-holes of a single site to $2$.
\emph{But, we want to introduce a description in terms of holes not in the way of change of variables but in the way
of a real change in the $d$-sector: where there are particles there should be holes and vice versa.}
\begin{figure}[H]
	\centering
	\includegraphics[scale=.25]{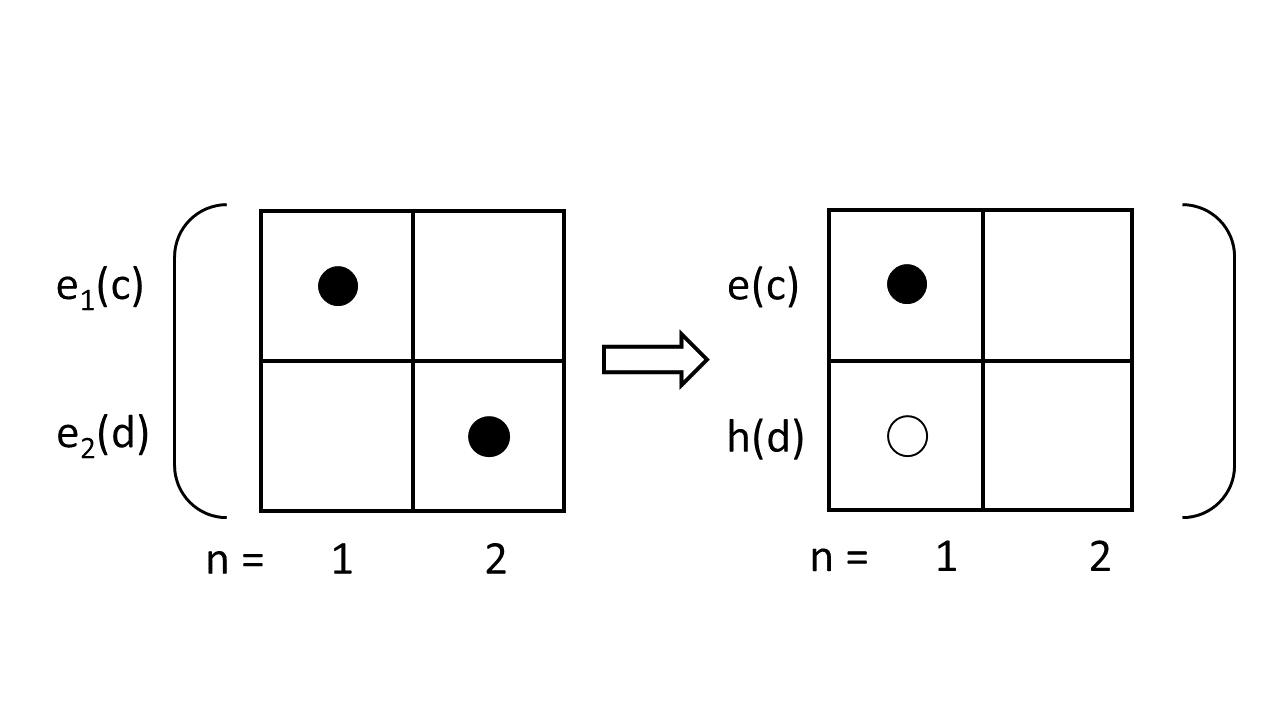}
	\caption{An illustration of the particle-hole transformation in the special bilayer problem. The transformation is done in the layer with particles (electrons) 2 and thus, also on the associated composite $d$-fermion.}
	\label{figure1}
\end{figure}

\begin{figure}[H]
	\centering
	\includegraphics[scale=.25]{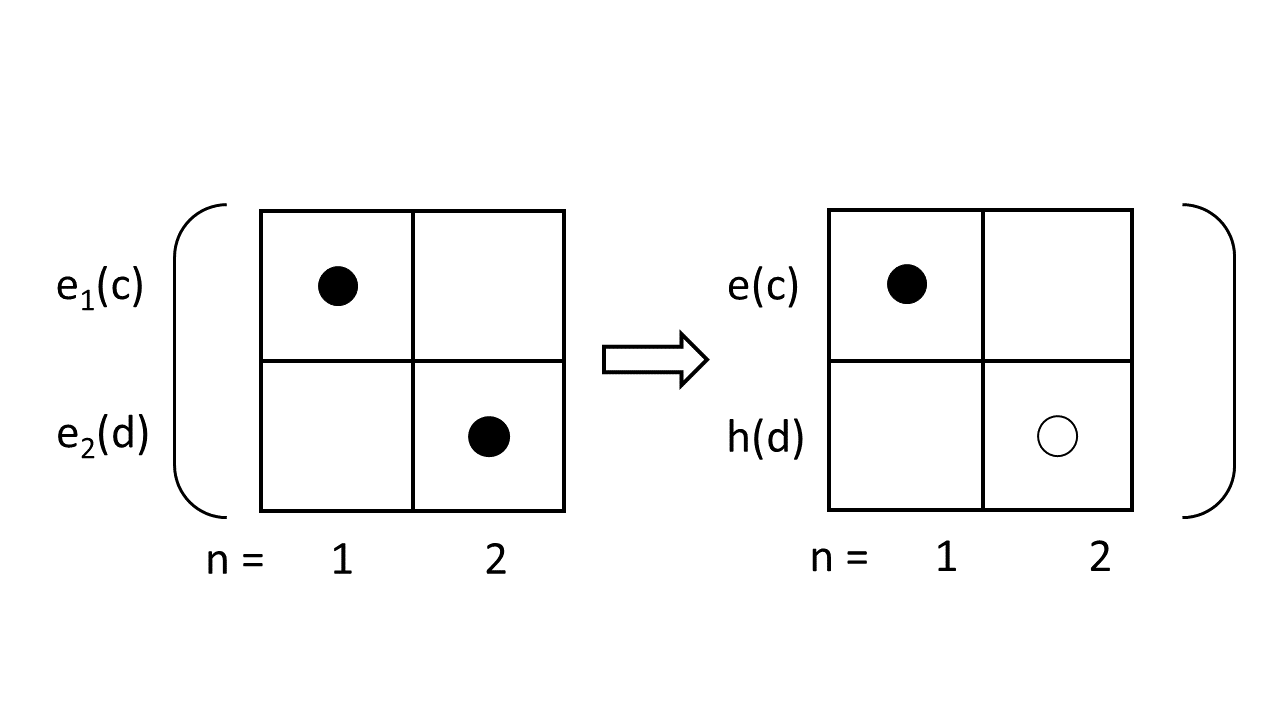}
	\caption{An illustration of the ``active" particle-hole transformation, i.e., the particle-hole conjugation on particles (electrons) 2 which become holes. In this way the special bilayer problem is transformed into the half-filled LL problem.}
	\label{figure2}
\end{figure}
Thus,
$\rho^{L(d)} \rightarrow \rho^{R(d)}$ and $\rho^{R(d)} \rightarrow \rho^{L(d)}$ in accordance with the discussion at the beginning of this section. In this way the operator
$\rho^{L(c)}_{n n} -\rho^{L(d)}_{n n}$ becomes $\rho^{L(c)}_{n n} -\rho^{R(d)}_{n n}$ and describes fluctuating charge of the half-filled LL. Also, the following action on operators $d_{m n}$ and $d_{m n}^\dagger $ is implied:
\begin{equation}
d_{m n} \rightarrow d_{n m}\;\; (d_{m n}^\dagger \rightarrow d_{n m}^\dagger ),
\label{trans}
\end{equation}
[compare the definitions of the density operators in (\ref{dend}) and (\ref{denrd}), in $L$ and $R$ sectors]. Fig. 1 and Fig. 2 illustrate the difference between the particle-hole transformation as a change of variables, and one that is an active transformation that transforms the special bilayer problem into the one of the half-filled LL.

\section{The formulation of the half-filled problem}

On the basis of the discussion in the previous section, we can conclude that the charge fluctuations around mean density $\frac{1}{2}(\frac{1}{2 \pi l_{B}^2})$, where $l_B$ is the magnetic length, can be expressed  by $ (\rho_{n n}^{L(c)} - \rho_{n n}^{R(d)})/2 $. Thus, the Hamiltonian that can be considered together with the hard-core constraints $\rho_{n n}^{R(c)} + \rho_{n n}^{L(d)} = 1$ and  $\rho_{n n}^{L(c)} + \rho_{n n}^{R(d)} = 1$ is
\begin{eqnarray}
{\cal H} &=&\frac{1}{2} \int d{\bs{q}} \;V(|{\bs{q}}|)   \label{phHam}\\
&& \times\frac{\rho^{L(c)}({\bs{q}})- \rho^{R(d)}({\bs{q}})}{2}
\frac{\rho^{L(c)}(-{\bs{q}})- \rho^{R(d)}(-{\bs{q}})}{2}. \nn
\end{eqnarray}
The charge operator $(\rho^{L(c)}({\bs{q}})- \rho^{R(d)}({\bs{q}}))/2$ does not satisfy the GMP
algebra because of the doubling of the
degrees of freedom (extra 2 in the GMP algebra). But together with the constraint $ \rho^{L(c)}({\bs{q}}) +\rho^{R(d)}({\bs{q}}) = 0$ it does, because $(\rho^{L(c)}({\bs{q}})- \rho^{R(d)}({\bs{q}}))/2$ becomes
$\rho^{L(c)}({\bs{q}})$ (or $-\rho^{R(d)}({\bs{q}})$ due to the PH symmetry) with the
constraint, and represents the physical charge that satisfies the GMP algebra, expressed as a change of the electron density $\rho^{L(c)}({\bs{q}})$ or a negative change in the hole density   $-\rho^{R(d)}({\bs{q}})$.

In the form of the Hamiltonian in (\ref{phHam}) we incorporated the PH symmetry, by including both particles and holes with equal weights (on an equal footing) in the description of the change of the physical (electric) charge from the mean value. The Hamiltonian is manifestly invariant under the transformation: $\rho^{L(c)} \rightarrow \rho^{R(d)}$ and $\rho^{R(d)} \rightarrow \rho^{L(c)} $, which represents an effective PH transformation. It consists of a charge conjugation, which transforms CFs, i.e., $c$'s into CHs, i.e., $d$'s, and vice versa, and an anti-unitary, i.e., time-reversal transformation that transforms phases into complex-conjugated ones $(R \rightarrow L $ and  $R \rightarrow L $). Also, the transformation  incorporates the reversal of momenta, $ {\bs{k}} \rightarrow - {\bs{k}}$, that can be associated with the time reversal, and thus, $\rho^{L(c)} ({\bs{q}}) \rightarrow \rho^{R(d)}(-{\bs{q}}) $ and $\rho^{R(d)}({\bs{q}}) \rightarrow \rho^{L(c)}(-{\bs{q}})$. The charge conjugation and time reversal constitute the usual definition of the PH transformation in the presence of the magnetic field, under which the physics should be invariant. We should note that the charge conjugation in our case is not uniquely defined on $c$'s and $d$'s and we get the same transformations of $\rho^{L(c)} $ and $ \rho^{R(d)} $ by considering $c_{{\bs{k}}} \rightarrow  \alpha d_{-{\bs{k}}} $ and $d_{{\bs{k}}} \rightarrow  \beta c_{-{\bs{k}}}$, where $\alpha$ and $\beta$ are constants, and $|\alpha| = |\beta| = 1$. 

We will recapitulate the necessary constraints in the formulation of the half-filled LL problem.
We summarize that
\begin{equation}
\rho_{n n}^{R(c)} + \rho_{n n}^{L(d)} = 1,
\label{r1}
\end{equation}
together with the definite statistics requirement in the unphysical sector,
and
\begin{equation}
\rho_{n n}^{L(c)} + \rho_{n n}^{R(d)} = 1,
\label{r2}
\end{equation}
with global constraints,
\begin{equation}
\sum_{n} \rho_{n n}^{L(c)} = \sum_{n} \rho_{n n}^{R(d)} = N/2,
\label{r3}
\end{equation}
and
\begin{equation}
\sum_{n} \rho_{n n}^{R(c)} = \sum_{n} \rho_{n n}^{L(d)} = N/2,
\label{r4}
\end{equation}
form a set of constraints that define the half-filled LL problem.

In this case physical states are
\begin{eqnarray}
&& \vert n_{1},\dots,n_{N/2},n'_{1},\dots,n'_{N/2}\rangle   \nonumber \\
&& =\sum\limits_{m_{1},\dots,m_{N/2}, m'_{1},\dots,m'_{N/2}}^{N} s^{m_{1} \cdots m_{N/2} m'_{1},\dots,m'_{N/2}}  \nonumber \\
&& \times c_{m_{1}n_{1}}^{\dagger} \cdots c_{m_{N/2} n_{N/2}}^\dagger d_{n'_{1} m'_{1}}^{\dagger} \cdots d_{n'_{N/2} m'_{N/2} }^\dagger \vert 0\rangle,
\label{phstates}
\end{eqnarray}
where $s^{m_{1} \cdots m_{N/2} m'_{1},\dots,m'_{N/2}}$ is non-zero, equal to one only if no index is equal to
any other index. Compare with (\ref{enstates}), and the discussion and (\ref{trans}) at the end of the previous section. Also, in (\ref{phstates}), $n_{i}\neq n'_{j}$ for any $i,j=1,\dots,N/2$.

\section{Preferred form of the Hamiltonian}

The most natural binding in ${\cal H}$ is the Cooper pair binding $\langle c_{\bs{k}} d_{-{\bs{k}}}\rangle \neq 0 $
in the $s$-wave channel. In a Hartree-Fock treatment, the mean-field description would have kinetic terms with quadratic dispersions, for $c$ and $d$ degrees of freedom, but this description of these objects does not conform to our expectation that they are dipoles - distinct dipole objects that pair, and that their dispersion comes from the polarization energy due to their dipole moments in a Hartree contribution as emphasized in \cite{dose}. In this way, we see a reason why the so-called PH Pfaffian, connected with $s$-wave pairing, is absent in a fixed LL \cite{avm,bal,mish,ym}.

As in the $\nu = 1$ bosonic case we may wonder whether there exists a ``preferred" form of the Hamiltonian, i.e.,
the Hamiltonian with some of constraints included in its formulation but with the same description (and
action) as the original one in the physical space. The ``preferred" form should capture the basic physics in the most efficient way, enabling the description of the basic physics in a Hartree-Fock treatment.

It is not hard to see
that a unique low-momentum possibility for a kinetic (non-pairing) term can be reached by an addition of the following term,
\begin{eqnarray}
{\cal H} &&\rightarrow {\cal H} + \frac{1}{2} \int d{\bs{q}} \;V(|{\bs{q}}|)   \\
&& \times \frac{\rho^{R(c)}({\bs{q}}) + \rho^{L(d)}({\bs{q}})}{2}
\frac{\rho^{R(c)}(-{\bs{q}}) + \rho^{L(d)}(-{\bs{q}})}{2}, \nonumber  \label{HPP}
\end{eqnarray}
which uses  the following constraint,
\begin{equation}
\rho^{R(c)}({\bs{q}}) + \rho^{L(d)}({\bs{q}}) = 0,
\end{equation}
in the physical sector for the unphysical degrees of freedom
that directly follows from the requirement (\ref{r1}).

In this way we removed the cause for the $s$-wave Cooper pairing and modified the relevant term from
\begin{equation}
\sim \int d{\bs{q}}\left[-\rho^{L(c)}({\bs{q}}) \rho^{R(d)}(-{\bs{q}})\right] V(|{\bs q}|)
\end{equation}
to
\begin{eqnarray}
&&\sim \int d{\bs{q}}\left[- \rho^{L(c)}({\bs{q}}) \rho^{R(d)}(-{\bs{q}}) + \rho^{R(c)}({\bs{q}}) \rho^{L(d)}(-{\bs{q}})\right] V(|{\bs q}|) \nn\\
&&\sim \int d{\bs{q}} \int d{\bs{k}}_1 \int d{\bs{k}}_2\; c_{{{\bs{k}}_1}- {\bs{q}}}^\dagger c_{{\bs{k}}_1}
d_{{{\bs{k}}_2}+{\bs{q}}}^\dagger d_{{\bs{k}}_2}\\
&&\hspace{4.5cm} \times \left[i ({\bs{k}}_1 + {\bs{k}}_2 ) \times {\bs{q}}\right] V(|{\bs q}|). \nonumber \label{expansion}
\end{eqnarray}
Clearly, this term in the Hartree-Fock treatment can lead only to a $\langle c_{\bs{k}}^\dagger d_{{\bs{k}}}\rangle \neq 0$ instability and a Dirac-type description of the low-momentum physics.

\section{The Dirac theory from the mean field}
Thus, we apply the Hartree-Fock approach to the relevant part of the Hamiltonian (we neglect the quadratic contributions from the other terms),
\begin{eqnarray}
&& {\cal H}_D = \int \frac{d{\bs{q}}}{4} V(|{\bs q}|)\Big[- \rho^{L(c)}({\bs{q}}) \rho^{R(d)}(-{\bs{q}}) + \rho^{R(c)}({\bs{q}}) \rho^{L(d)}(-{\bs{q}})\Big]    \nonumber \\
&& \approx\int d{\bs{q}} \int d{\bs{k}}_1 \int d{\bs{k}}_2  \;\frac{V(|{\bs q}|)}{4 (2 \pi)^4 } [i ({\bs{k}}_1 + {\bs{k}}_2 ) \times {\bs{q}} ] \nonumber \\
&&   \times\Big[ \langle c_{{{\bs{k}}_1}- {\bs{q}}}^\dagger d_{{\bs{k}}_2}\rangle
d_{{{\bs{k}}_2}+{\bs{q}}}^\dagger c_{{\bs{k}}_1}  \nonumber \\
&&+c_{{{\bs{k}}_1}- {\bs{q}}}^\dagger d_{{\bs{k}}_2}
\langle d_{{{\bs{k}}_2}+{\bs{q}}}^\dagger c_{{\bs{k}}_1}\rangle -
\langle c_{{{\bs{k}}_1}- {\bs{q}}}^\dagger d_{{\bs{k}}_2}\rangle
\langle d_{{{\bs{k}}_2}+{\bs{q}}}^\dagger c_{{\bs{k}}_1}\rangle\Big] \nonumber \\
&& =\int \frac{d{\bs k}}{(2 \pi)^2} \left(\Delta_{\bs k}^{*} d_{\bs k}^\dagger c_{\bs k} + \Delta_{\bs k} c_{\bs k}^\dagger d_{\bs k}\right) + {\cal C},
\end{eqnarray}
where ${\cal C}$ is a constant and
\begin{equation}
\Delta_{\bs k} = |{\bs k}| \int d{\bs q}\; \frac{V(|{\bs q}|)}{2 (2 \pi)^2} (i{\hat{\bs{k}}} \times {\bs q})
\langle d_{{\bs k}+{\bs q}}^\dagger c_{{\bs k}+{\bs q}}\rangle.
\label{dedef}
\end{equation}
We diagonalize ${\cal H}_D $ by introducing $\alpha_{\bs k}$ and $\beta_{\bs k}$ operators,

\begin{equation}
\left[\begin{array}{c}  c_{\bs k} \\
                                      d_{\bf k} \\

\end{array} \right]= \frac{1}{\sqrt{2}}
\left[\begin{array}{cc}  1 & - \exp\{- i \delta_{\bs k}\}\\
                              \exp\{i \delta_{\bs k}\} & 1 \\

\end{array} \right]
\left[\begin{array}{c}  \alpha_{\bs k} \\
                                      \beta_{\bf k} \\

\end{array} \right],
\end{equation}
where $\delta_{\bf k}$ is defined by $ \Delta_{\bf k} = |\Delta_{\bf k}| \exp\{- i \delta_{\bf k}\}$.

The very important question is how we choose the occupation of the momentum ${\bs k}$ states in the
ground state that is subjected to the constraints,
\begin{equation}
\rho_{n n'}^{R(c)} + \rho_{n' n}^{L(d)} = \delta_{n n'},
\label{c1}
\end{equation}
and
\begin{equation}
\rho_{n n'}^{L(c)} + \rho_{n' n}^{R(d)} = \delta_{n n'}.
\label{c2}
\end{equation}
In a mean field treatment we expect that at least the global constraints,
\begin{equation}
\int \frac{d{\bs k}}{(2\pi)^2 } c^\dagger_{\bs k} c_{\bs k} = \int \frac{d{\bs k}}{(2\pi)^2 } d^\dagger_{\bs k}d_{\bs k} = \bar{\rho}_e = \frac{1}{2} \frac{1}{2\pi l_B^2 }, \label{cdc}
\end{equation}
will be satisfied.

In the $\alpha_{\bs k}$, $\beta_{\bs k}$ language this implies
\begin{equation}
\int d{\bs k} \left(e^{-i \delta_{\bs k}} \alpha_{\bs k}^\dagger \beta_{\bs k} + e^{i \delta_{\bs k}} \beta_{\bs k}^\dagger \alpha_{\bs k}\right) = 0.
\label{glocd}
\end{equation}
This is a complex constraint and we may try to satisfy the requirement on $\alpha$'s and $\beta$'s, by demanding that also the number of $\alpha$'s and $\beta$'s is conserved. We might expect,
\begin{equation}
\int \frac{d{\bs k}}{(2\pi)^2 } \alpha^\dagger_{\bs k} \alpha_{\bs k} = \int \frac{d{\bs k}}{(2\pi)^2 } \beta^\dagger_{\bs k}\beta_{\bs k} = \bar{\rho}_e.
\label{gloab}
\end{equation}
This seems a very crude ``translation" of (\ref{glocd}), but it incorporates the basic idea of our approach: to treat the particles and holes in an equal way, with their dynamics not independent but constrained, and in
this way duplicated in a theory. The constraint implies two sectors, $\alpha$ and $\beta$, in the ground-state configuration: half-filled $\alpha$-sector and half-empty $\beta$-sector.
\begin{figure}[H]
	\centering
	\includegraphics[scale=.3]{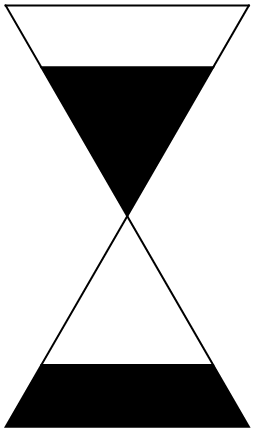}
	\caption{A schematic illustration of the implementation of the global constraint in (\ref{cdc}) via (\ref{gloab}), i.e., half-filled positive energy sector and half-empty negative energy sector.}
	\label{figure3}
\end{figure}
In this we implicitly assumed the finiteness of the available volume of ${\bs k}$: the number of available ${\bs k}$'s is $N$, the number of orbitals in the fixed LL. We expect that the description is duplicated by treating particles and holes in an equal way, and, in the first (mean-field) approximation, the dynamics of $\alpha$ and $\beta$ are separate and independent, and we may consider one or the other sector as a description of the problem.

Thus, for $\Delta_{\bs k}$ we get, by self-consistency, 
\begin{equation}
\Delta_{\bs k}  = |{\bs k}| e^{- i \phi_{\bs k}} \int_{|{\bs q}| \leq q_F} d{\bs q} \frac{V(|{\bs q}|)}{4 (2 \pi )^2}
( i \hat{\bs{ k}} \times {\bs q}) e^{- i\phi_{{\bs k}  + {\bs q}} + i \phi_{\bs k}},
\end{equation}
where $q_F = 1/l_{B} (=1)$.
In this expression for $\Delta_{\bs k}$, because of the Gaussian in $V(|{\bs q}|)$, and the long-wavelength, $|{\bs k}| \sim 0$, approximation, the contribution of the $\alpha$
sector is taken into account and we neglected the contribution from the $\beta$ sector. We choose $\delta_{\bs k}$ to
describe a definite momentum state, $\delta_{\bs k} = \phi_{\bs k}$, where $\phi_{\bs k}$ is the phase of the complex variable,
$ k = k_x + i k_y $. It follows that
\begin{equation}
|\Delta_{\bs k}| = |{\bs k}| \frac{\pi}{4 (2 \pi)^2} \int_0^{q_F} dq \; q^2 V(q).
\end{equation}
The strength of the amplitude is zero for higher angular momenta (other than angular momentum one).

Thus, by applying the Hartree-Fock approach to the preferred form of the Hamiltonian, (\ref{HPP}), we reached
a low-energy description of the problem in terms of
\begin{equation}
{\cal H}_D = \int \frac{d{\bs k}}{(2 \pi )^2} \left(\Delta_{\bs k}^{*} d_{\bs k}^\dagger c_{\bs k} + \Delta_{\bs k} c_{\bs k}^\dagger d_{\bs k}\right),
\label{df}
\end{equation}
where $ \Delta_{\bs k} = (k_x - i k_y ) \Delta $ with  $\Delta = \frac{\pi}{4 (2 \pi )^2} \int_0^{q_F} dq \; q^2 V(q)$, at the finite density of the Dirac system.

The system that is described by the Dirac Hamiltonian, at finite density, represents a Fermi liquid. The time reversal in this system transforms a state at momentum $ {\bs k}$ into a one with momentum $ -  {\bs k}$. Concretely, the state with momentum $ {\bs k}$:
\begin{equation}
\frac{1}{\sqrt{2}}  \left[
  \begin{array}{c}
   \exp\{-i \phi_{\bs k} \}  \\
   1 \\
  \end{array}
\right] ,
\end{equation}
where  $ \phi_{\bs k}$ is the phase of $ k_+ = k_x + i k_y $, is transformed, under the time reversal transformation, $U = - i \sigma_y K $, where $\sigma_y$ is the Pauli matrix, and $K$ denotes the complex conjugation, into
 \begin{equation}
\frac{1}{\sqrt{2}}  \left[
  \begin{array}{c}
   -1 \\
   \exp\{i \phi_{\bs k} \} \\
  \end{array}
\right],
\end{equation}
i.e., the state of the same energy but opposite momentum, $-{\bs k}$. In the language of the basic PH transformation on the system (that acts on elementary particles, electrons and holes), more precisely its realization on the particular description of the problem that we introduced, the time reversal transformation, in the Dirac description, corresponds to the following  transformation on $c$'s and $d$'s, $c_{{\bs{k}}} \rightarrow  - d_{-{\bs{k}}} $ and   $d_{{\bs{k}}} \rightarrow   c_{-{\bs{k}}} $
(or  $c_{{\bs{k}}} \rightarrow   d_{-{\bs{k}}} $ and  $d_{{\bs{k}}} \rightarrow -  c_{-{\bs{k}}} $, compare with the description in Section V). 


\section{The inclusion of the gauge invariance in the effective Dirac theory}

The original $SU(N)$ gauge invariance (i.e., invariance under a change of basis in the fixed LL) is broken down \cite{dose} to $U(1)$ in the mean-field (Hartree-Fock, averaged) description in (\ref{df}).
As we already detailed, in the microscopic approach the $SU(N)$ gauge invariance is realized by the following two constraints,
\begin{equation}
\rho_{n n}^{R(c)} + \rho_{n n}^{L(d)} = 1,
\label{r11}
\end{equation}
i.e., equal charge distribution of unphysical degrees of freedom,
and
\begin{equation}
\rho_{n n}^{L(c)} + \rho_{n n}^{R(d)} = 1,
\label{r22}
\end{equation}
i.e., the exclusion of the double occupancy between particles and holes (extra unphysical degrees of freedom).


To include fluctuations beyond the Hartree-Fock (mean-field) level in the long-wavelength domain, Dong and Senthil reinstated the $SU(N)$ invariance in the description of the boson problem in terms of CFs, by introducing composite fermion fields on noncommutative (deformed) space,
\begin{equation}
c({\bs R},\tau) = \int \frac{d^{2}\bs{k}}{(2 \pi)^{\frac{3}{2}}} \exp\left(i \bs{k} \cdot\bs{R}\right)  c_{\bs{k},\tau},
\end{equation}
where ${\bs R}$ is the noncommutative (guiding center) coordinate, and $\tau$ is imaginary time. The connection with the microscopic description is the following: 
\begin{equation}
c_{n m} =  \langle n\vert c({\bs R},\tau)\vert m\rangle =  \int \frac{d^{2}\bs{k}}{(2 \pi)^{\frac{3}{2}}} \langle n\vert\tau_{\bs{k}}\vert m\rangle c_{\bs{k},\tau}.
\end{equation}
Physically, the state of the CF with vortex orbital $m$ and electron 
orbital $n$, can be described by a superposition of the (commutative) 
momentum ${\bs k}$ states, the weights of which depend on the effective 
distance between orbitals (the size of the dipole), $|{\bs k}_{\rm 
eff}|$, because 
 $ \tau_{\bs{k}} = \exp\left(i \bs{k} \cdot\bs{R}
\right)$ is the translation operator.  For more details see \cite{dose}. In this way, the $L$ and $R$ transformations on $c_{n m}
$,
\begin{equation}
c_{nm}\rightarrow U^{L}_{n' n} c_{n m} U^{R}_{m m'},
\end{equation}
can be represented by noncommutative Moyal-Weyl star-product of fields on ordinary commutative space,
\begin{equation}
c({\bs x},\tau)\rightarrow U^{L}({\bs x},\tau)  \star c({\bs x},\tau) \star U^{R}({\bs x},\tau) ,
\end{equation}
the star-product being defined by 
\begin{equation}
a({\bs x},\tau) \star  b({\bs y},\tau)=e^{-\frac{i}{2}l^{2}_{B}\epsilon^{ij}\frac{\partial}{\partial x^{i}}\frac{\partial}{\partial y^{j}}} a({\bs x},\tau) b({\bs y},\tau)\vert_{{\bs y}\rightarrow {\bs x}}.
\end{equation}
Two-dimensional Levi-Civita ($i,j=1,2$) is defined by $\epsilon^{12}=-\epsilon^{21}=1$. See Appendix A for a more elaborate account.

The two constraints in our problem, (\ref{r11}) and (\ref{r22}), imply, in the effective description, simultaneous $SU(N)$ transformations in $L$ and $R$ sectors to which we may associate, gauge fields $a_\mu^{(1)}$ and $a_\mu^{(2)}$. We may also consider a (background) field $A_\mu$ that is associated with physical degrees of freedom (particles and holes), $\rho_{n n}^{L(c)} - \rho_{n n}^{R(d)}$.

Thus, we expect the corresponding covariant derivatives of the following form,
\begin{align}
D_\mu c &= \partial_\mu c - i (a_\mu^{(1)}+A_\mu) \star c- i c \star a_\mu^{(2)}, \label{DDc}\\
D_\mu d &= \partial_\mu d - i a_\mu^{(2)} \star d - i  d \star (a_\mu^{(1)}-A_\mu), \label{DDd}
\end{align}
in the noncommutative description of the low-energy physics, following the considerations in
\cite{dose} for the $\nu = 1$ system of bosons.

But there is a problem with this proposal for a noncommutative description: the structure of these derivatives is not consistent with unitarity of gauge transformations, see the Appendix B for details. 


We have to step back to understand why this problem occurs. The formulation of the half-filled LL with constraints (\ref{r11}) and (\ref{r22}) is different from the case of bosons at $\nu = 1$ filling, in which there is a clear distinction between $L$ and $R$, physical and unphysical sector, 
The formulation of the half-filled LL system 
is more intricate since it includes an exchange of $L$ and $R$ sectors, which complicates the distinction between them. 

Having this in mind, we may reconsider the question of constraints and gauge invariance in an effective, long-wavelength theory that we are looking for, a theory that will nevertheless include some noncommutative aspects of the physical system. In order to get a gauge-invariant description, we will also apply the long-wavelength limit on the constraints (not just in the derivation of the Hamiltonian).   
In the ensuing long-wavelength description there is no distinction between physical (particles or holes) and unphysical (quantum of flux excitation) degrees of freedom,  constraints (\ref{r11}) and (\ref{r22}) become one [compare (\ref{bden}) and (\ref{bdenr}) in the small-$q$ limit, etc.]. We may also expect that $A_\mu $ (background field) couples symmetrically to the unphysical and physical sector. 
Thus, we may define covariant derivatives $D_\mu c$ and $D_\mu d $ in a way that ensures gauge invariance of the theory,
\begin{align}
D_\mu c &= \partial_\mu c - i  A_\mu \star c  - i c\star (a_\mu-A_\mu ),\label{Dc}\\
D_\mu d &= \partial_\mu d - i  A_\mu \star d  - i d\star (a_\mu-A_\mu ).\label{Dd}
\end{align}
The $R$ gauge field is introduced in a decomposed way, $a_{\mu}-A_{\mu}$, that will be natural in the commutative limit.

The change that we introduced by going from (\ref{DDc}) and (\ref{DDd}) to (\ref{Dc}) and (\ref{Dd}) is certainly drastic; the change is a departure from the microscopic description that we found based on the view of quasiparticles as neutral composites. In the case of bosons at filling factor one, Dong and Senthil showed that, on the basis of the Hartree-Fock description that is modified to include the $SU(N)$ invariance, in the manner of noncommutative field theory, an approximate commutative field theory description can be reached (via Seiberg-Witten map) in the form of Halperin-Lee-Read (HLR) \cite{hlr}  description with a Chern-Simons (CS) term. This description cannot be viewed as a HLR theory (although of the same form), but as one that is based on the neutral composites (CFs) and limited to a LL, a description that is analogous to the description by composite bosons of the Laughlin case in \cite{rcs}. 
In the case of electrons, in the half-filled LL, for which the Son's theory is relevant, one expects that, due to the requirement for the PH symmetry, a microscopic description will not generate a Chern-Simons term in an effective description. We encountered difficulties in the application of the program proposed by Dong and Senthil - to maintain the gauge invariance we consider constraints in the long-wavelength limit, which certainly implies changes in the microscopic physics, i.e., ultra-violet domain. We will find that the ensuing field theory will have a Chern-Simons term, but one may argue that now Pauli-Villars type of regularization is associated with the field theory, i.e., we have to treat the high-energy physics in a different way as opposed to the version of the Son's theory that was firstly proposed by Son, and assumes the dimensional regularization. Therefore, although we have done a drastic change in the microscopic domain, the theory may still make sense as a theory based on another kind of quasiparticle \cite{avm,mdva} and give a version of the Son's theory described in \cite{pot,wang,ssww}.

The covariant derivatives (\ref{Dc}) and (\ref{Dd}) can define an NC description, and it should be checked whether in the commutative limit via the Seiberg-Witten map we can recover the Son's theory to the linear order in the small parameter $\theta=-l_B^2 $. Because of the simultaneous presence of small-$\theta$ and long-wavelength expansions, we will seek an effective description by considering only lowest order terms. To find the first correction to the commutative limit we start with the (Euclidean) NC action of the form:
\begin{align}\label{NC_action}
&S_{NC}=\int d\tau d^{2}\boldsymbol{x}\;\Big(c^{\dagger} \star D_\tau c + d^{\dagger} \star D_\tau d \\
&+c^\dagger \star (iD_x +D_y)d 
+ d^\dagger \star (iD_x-D_y)c 
+ i (a_{0}-A_0)\bar{\rho}_e \Big).\nn
\end{align}

In $S_{NC}$ we have a constraint term, linear in NC field $a_0 - A_0$, that fixes the total number of $c$ and $d$ fermions. Recall that our description is for $0 \leq |{\bs k}|$ in the upper half of Fig. 3, and thus, we have  
\begin{equation}
\int \frac{d^{2}{\bs k}}{(2\pi)^2 }\left(c^\dagger_{\bs k} c_{\bs k} + d^\dagger_{\bs k}d_{\bs k}\right)|_{\text{upper  half}} = \bar{\rho}_e .
\end{equation}

In Appendix A we detail the small-$\theta$ expansion in terms of commutative fields $\hat{c} , \hat{d} , \hat{a}_\mu , \hat{A}_\mu $ (denoted by a hat symbol). We find $S_{NC}=S^{(0)}+S^{(1)}+\dots$,
where the classical limit ($\theta=0$) is simply
\begin{align}\label{Com_action}
S^{(0)}&=\int d\tau d^{2}\boldsymbol{x}\;\Big( \hat{c}^{\dagger}  D_\tau \hat{c} + \hat{d}^{\dagger} D_\tau \hat{d} \\
&+\hat{c}^\dagger  (iD_x +D_y)\hat{d} 
+ \hat{d}^\dagger (iD_x-D_y)\hat{c} 
+ i (\hat{a}_{0}-\hat{A}_0) \bar{\rho}_e \Big),\nn
\end{align}
and the linear NC correction reads as (see Appendix A for details):
\begin{widetext}
\begin{align}\label{NC_correction}
S^{(1)}=&\frac{i\bar{\rho}_{e}\theta}{2}\epsilon^{\alpha\beta\gamma}\int d\tau d^{2}\boldsymbol{x}\;(\hat{a}_{\alpha}-\hat{A}_{\alpha})\partial_{\beta}(\hat{a}_{\gamma}-\hat{A}_{\gamma})\nonumber\\
+&\theta\int{\rm d}\tau {\rm d}^{2}\boldsymbol{x}\left[\hat{c}^{\dagger}\left(\frac{1}{2}\hat{f}_{12}-\hat{F}_{12}\right)D_{\tau}\hat{c}-\hat{c}^{\dagger}\left(\frac{1}{2}\hat{f}_{10}-\hat{F}_{10}\right)D_{y}\hat{c}+\hat{c}^{\dagger}\left(\frac{1}{2}\hat{f}_{20}-\hat{F}_{20}\right)D_{x}\hat{c}\right]\nonumber\\
+&\theta\int{\rm d}\tau {\rm d}^{2}\boldsymbol{x}\left[\hat{d}^{\dagger}\left(\frac{1}{2}\hat{f}_{12}-\hat{F}_{12}\right)D_{\tau}\hat{d}-\hat{d}^{\dagger}\left(\frac{1}{2}\hat{f}_{10}-\hat{F}_{10}\right)D_{y}\hat{d}+\hat{d}^{\dagger}\left(\frac{1}{2}\hat{f}_{20}-\hat{F}_{20}\right)D_{x}\hat{d}\right],
\end{align}
\end{widetext}
where we introduced classical (commutative) gauge field strengths,
\begin{align}
\hat{F}_{\mu\nu}&=\partial_{\mu}\hat{A}_{\nu}-\partial_{\nu}\hat{A}_{\mu},\\
\hat{f}_{\mu\nu}&=\partial_{\mu}\hat{a}_{\nu}-\partial_{\nu}\hat{a}_{\mu},
\end{align}
and $\varepsilon^{\alpha\beta\gamma}$ is the Levi-Civita symbol.

The classical action $S^{(0)}$ and the Chern-Simons term in $S^{(1)}$ give the description of a version of the Son's theory \cite{son}
of the Dirac composite fermion \cite{pot,wang,ssww} that assumes the Pauli-Villars type of regularization \cite{avm,mdva} because
of the presence of the CS term for field $a_\mu $. Note that the CS term has the correct coefficient, $\tfrac{1}{8\pi}$. Also, in the linear NC correction $S^{(1)}$ the Dirac momentum density couples to the external electric field as expected from the Galilean invariance \cite{ph,pr}. On the other hand, the presence of a coupling to the internal electric field $\hat{a}_\mu$ is quite natural and expected given the influence of other particles on a selected one. Also, we find that the presence of internal and external (i.e., departure
from the uniform) magnetic field induces a change in the coefficient of the kinetic terms $\hat{c}^\dagger \partial_\tau \hat{c}$ and $\hat{d}^\dagger \partial_\tau \hat{d}$. Thus, we can conclude that the NC formulation, up to the first order in $\theta$, recovers known results but also systematically adds terms that we may expect on physical grounds.

Although we reproduced the version of the Son's theory from a conjectured NC field theory that is partially based on the microscopic approach, we may wonder whether there is a formulation of an NC field theory that can reproduce, via the Seiberg-Witten map, the original version of the Son's theory. In that case, there would be no terms that are not invariant under the PH transformation, like the CS term for gauge field $a_{\mu}$, and we would not have to assume that their effect will be erased or canceled by a particular way of regularization.

With the experience of the previous derivation that resulted in Eqs. (\ref{Dc}) and (\ref{Dd}) for covariant derivatives [and action \ref{NC_action}], we may ask ourselves how Eqs. (\ref{DDc}) and (\ref{DDd}) can be modified in a way that they still express the microscopic constraints, but represent valid covariant derivatives. A way to do that is by including a symmetry between physical and unphysical sectors, by assuming that the background field $A_{\mu}$ couples to both sectors. 

Thus, we consider
\begin{align}
D_{\mu}c&=\partial_{\mu}c-i(a_{\mu}^{(1)}+A_{\mu})\star c-ic\star(a_{\mu}^{(2)}-A_{\mu}),\label{Dc_new} \\
D_{\mu}d&=\partial_{\mu}d-i(a_{\mu}^{(2)}+A_{\mu})\star d-id\star(a_{\mu}^{(1)}-A_{\mu}).\label{Dd_new}
\end{align}
Note that, as before, in the course of the implementing microscopic constraints, (\ref{r11}) and (\ref{r22}), where each one constrains densities both in $R$ and $L$ sector, the implied gauge fields connect i.e., transform, at the same time, in the $R$ and $L$ sectors of fields $c$ and $d$. We warn the reader that, as it stands, in the proposed covariant derivatives, only fields $b^{(1)}_{\mu,\pm}\equiv a_{\mu}^{(1)}\pm A_{\mu}$ and $b^{(2)}_{\mu,\pm}\equiv a_{\mu}^{(2)}\pm A_{\mu}$ are assumed to transform canonically (see Appendix B for details), and  fields  $a_{\mu}^{(1)}$,  $a_{\mu}^{(2)}$, and $ A_{\mu}$, that enter their decompositions (as they refer both to $R$ and $L$ sectors) do not. The complete set-up of the new NC theory is given in terms of $b^{(1)}_{\mu,\pm}$ and $b^{(2)}_{\mu,\pm}$ only. In the Appendix B we explain why necessarily $b^{(1)}_{\mu,\pm} = b^{(2)}_{\mu,\pm}$ = $b_{\mu,\pm}$, i.e., $a_{\mu}^{(1)} = a_{\mu}^{(2)}$, and we have an emerging symmetry between the unphysical and physical sectors. The fields $a_{\mu} = b_{\mu,+} + b_{\mu,-}$  and $A_{\mu} = (b_{\mu,+} - b_{\mu,-})/2 $ will assume their expected $U(1)$ gauge field roles in the commutative limit of the new theory in which we get 
the original version of the Son's theory that incorporates the PH symmetry (CP or CT in \cite{son}). The resulting NC action up to first order is given by
\begin{widetext}
\begin{align}
S_{NC}&=\int d\tau d^{2}\boldsymbol{x}\Big( \hat{c}^{\dagger}  D_\tau \hat{c} + \hat{d}^{\dagger} D_\tau \hat{d} 
+\hat{c}^\dagger  (iD_x +D_y)\hat{d} 
+ \hat{d}^\dagger (iD_x-D_y)\hat{c} 
+ i \hat{a}_{0} \bar{\rho}_e \Big)+\frac{i}{4\pi}\varepsilon^{\alpha\beta\gamma}\int d\tau d^{2}\boldsymbol{x}\;A_{\alpha}\partial_{\beta}a_{\gamma}\nn\\
&+\theta\int d\tau d^{2}\boldsymbol{x}\left(\hat{c}^{\dagger}\hat{F}_{12}D_{\tau}\hat{c}-\hat{c}^{\dagger}\hat{F}_{10}D_{y}\hat{c}+\hat{c}^{\dagger}\hat{F}_{20}D_{x}\hat{c} 
+\hat{d}^{\dagger}\hat{F}_{12}D_{\tau}\hat{d}-\hat{d}^{\dagger}\hat{F}_{10}D_{y}\hat{d}+\hat{d}^{\dagger}\hat{F}_{20}D_{x}\hat{d} \right).\label{NC_Son}
\end{align}
\end{widetext}

Thus, we obtained an effective (long-wavelength) NC field theory that consistently reproduces the Son's theory in the commutative limit. The introduced formalism and NC setup can be used to systematically generate corrections in small parameter $\theta$.
The achieved NC descriptions need further understanding and analysis especially concerning the questions of regularization and scaling of gauge fields \cite{son,ssww,knp}. At first glance, it seems that our approach does not have the scaling problem.

\section{Inclusion of pairing}

We have eliminated unphysical degrees of freedom in a way of constraint (\ref{r1}), although in the case of the
half-filled Landau level we also had to impose additional bosonic correlations of the unphysical degrees of freedom to fix a unique subspace of physical states - we called it spin-singlet sector of the $SU(N)$ gauge symmetry of unphysical degrees of freedom. [Subsequently, we also had to impose (\ref{r2}) - to eliminate
hole degrees of freedom.]

We may search for another such state for unphysical degrees of freedom by imposing other constraint(s), such as
\begin{equation}
\rho_{n n'}^{R(c)} = \rho_{n' n}^{L(d)}.
\label{c3}
\end{equation}
In the inverse space this corresponds to
\begin{equation}
\rho^{R(c)}({\bs{q}}) - \rho^{L(d)}({\bs{q}}) = 0.
\label{natreq}
\end{equation}
This  choice seems natural as a requirement that will equalize and uniformly distribute the electric charge of the unphysical degrees of freedom (similarly to the special bilayer system).

The Hamiltonian that we can consider now is
\begin{eqnarray}
{\cal H}_p && = {\cal H} - \frac{1}{2} \int d^{2}{\bs{q}}\; V(|{\bs{q}}|)   \\
&& \times\frac{\rho^{R(c)}({\bs{q}}) - \rho^{L(d)}({\bs{q}})}{2}
\frac{\rho^{R(c)}(-{\bs{q}}) - \rho^{L(d)}(-{\bs{q}})}{2},\nn \label{HP}
\end{eqnarray}
which contains the same relevant two-body part for the Dirac physics as in the previous inclusion of constraints (\ref{HPP}), but now we have the diagonal terms in the $c$ and $d$-sector, like
\begin{eqnarray}
&&\sim \int d^{2}{\bs{q}} \left[\rho^{L(c)}({\bs{q}}) \rho^{L(c)}(-{\bs{q}}) - \rho^{R(c)}({\bs{q}}) \rho^{R(c)}(-{\bs{q}})\right] V(|{\bs q}|) \nonumber \\
&&\sim \int d^{2}{\bs{q}} \int d^{2}{\bs{k}}_1 \int d^{2}{\bs{k}}_2\; c_{{{\bs{k}}_1}- {\bs{q}}}^\dagger c_{{\bs{k}}_1}
c_{{{\bs{k}}_2}+{\bs{q}}}^\dagger c_{{\bs{k}}_2}\\ &&\hspace{4.5cm} \times i\left[({\bs{k}}_1 - {\bs{k}}_2 ) \times {\bs{q}}\right] V(|{\bs q}|), \nn
\end{eqnarray}
in the $c$-sector, that can lead to $p$-wave (Pfaffian in the $c$-sector and anti-Pfaffian in the $d$-sector) instabilities.

We will assume a PH symmetry breaking and an effective Hartree-Fock-BCS Hamiltonian of the
following form,
\begin{eqnarray}
{\cal H}_{BCS} &=& \int \frac{d^{2}{\bs k}}{(2 \pi )^2} \left(\Delta_{\bs k}^{*} d_{\bs k}^\dagger c_{\bs k} + \Delta_{\bs k} c_{\bs k}^\dagger d_{\bs k}\right) \nonumber \\
+&&  \int \frac{d^{2}{\bs k}}{(2 \pi )^2} \left({\tilde\Delta}_{\bs k}^{*} c_{-\bs k} c_{\bs k} + {\tilde\Delta}_{\bs k} c_{\bs k}^\dagger c_{-\bs k}^\dagger\right).
\end{eqnarray}
Here, $\Delta_{\bs k}$ is defined in (\ref{dedef}) and
\begin{equation}
{\tilde\Delta}_{\bs k}^{*} = |{\bs k}| \int d^{2}{\bs q}\; \frac{V(|{\bs q}|)}{4 (2 \pi)^2} (i\hat{\bs{k}} \times {\bs q})
\langle c_{{\bs k}-{\bs q}}^\dagger c_{-{\bs k}+{\bs q}}^\dagger \rangle.
\label{tdedef}
\end{equation}
We project to the $\alpha$ sector by taking
\begin{equation}
c_{\bs k} \rightarrow \frac{1}{\sqrt{2}} \; \alpha_{\bs k},
\end{equation}
and
\begin{equation}
d_{\bs k} \rightarrow \frac{1}{\sqrt{2}} e^{i \phi_{\bs k}}  \alpha_{\bs k}.
\end{equation}
Thus,
\begin{equation}
{\cal H}_{BCS}^\alpha = \int \frac{d^{2}{\bs k}}{(2 \pi )^2} \left[|\Delta_{\bs k}| \alpha_{\bs k}^\dagger \alpha_{\bs k} + \left(\frac{{\tilde\Delta}_{\bs k}^{*}}{2} \alpha_{-\bs k} \alpha_{\bs k} + h.c.\right)\right].
\end{equation}
Equations \cite{rg} that follow and need to be solved self-consistently are
\begin{equation}
|\Delta_{\bs k}| = |{\bs k}| \int d^{2}{\bs q}\; \frac{V(|{\bs q}|)}{8 (2 \pi)^2}  |{\bs q}| \sin^2 (\phi_{\bs  q}) \left(1 - \frac{|\Delta_{{\bs k} + {\bs q}}| - |\Delta_{q_F}|}{E_{{\bs k} + {\bs q}}}\right),
\label{sc1}
\end{equation}
and
\begin{equation}
|{\tilde \Delta}_{\bs k}| = |{\bs k}| \int d^{2}{\bs q}\; \frac{V(|{\bs q}|)}{16 (2 \pi)^2}  |{\bs q}| \sin^2 (\phi_{\bs  q})  \frac{|{\tilde \Delta}_{{\bs k} + {\bs q}}|}{E_{{\bs k} + {\bs q}}},
\label{sc2}
\end{equation}
where $E_{\bs q}^2 = (|\Delta_{{\bs q}}| - |\Delta_{q_F}|)^2 + |{\tilde \Delta}_{\bs q}|^2$. In this way,
by specifying $V(|{\bs q}|)$ that will include factors due to the projection to a fixed Landau level, we can find amplitudes in $ {\tilde \Delta}_{\bs k} = | {\tilde \Delta}_{\bs k}| \exp\{- i \phi_{\bs k}\}$ and
$ {\Delta}_{\bs k} = | {\Delta}_{\bs k}| \exp\{-i \phi_{\bs k}\}$.

\section{Pairing solutions}

The effective interaction in a fixed LL, $ m = 0, 1, 2, \ldots$, is given by the following expression:
\begin{equation}
V(|{\bs q}|) = V_c (|{\bs q}|) e^{- \frac{|{\bs q}|^2}{2}} \left[L_m \left(\frac{|{\bs q}|^2}{2}\right)\right]^2,
\end{equation}
where
\begin{equation}
V_c (|{\bs q}|) = \frac{V_0}{|{\bs q}|},
\end{equation}
represents the Coulomb interaction, $l_B = 1$, and $L_m $ denotes the Laguerre polynomial associated with
a fixed LL with the quantum number $m$.

\begin{figure}
	\centering
	\includegraphics[scale=.18]{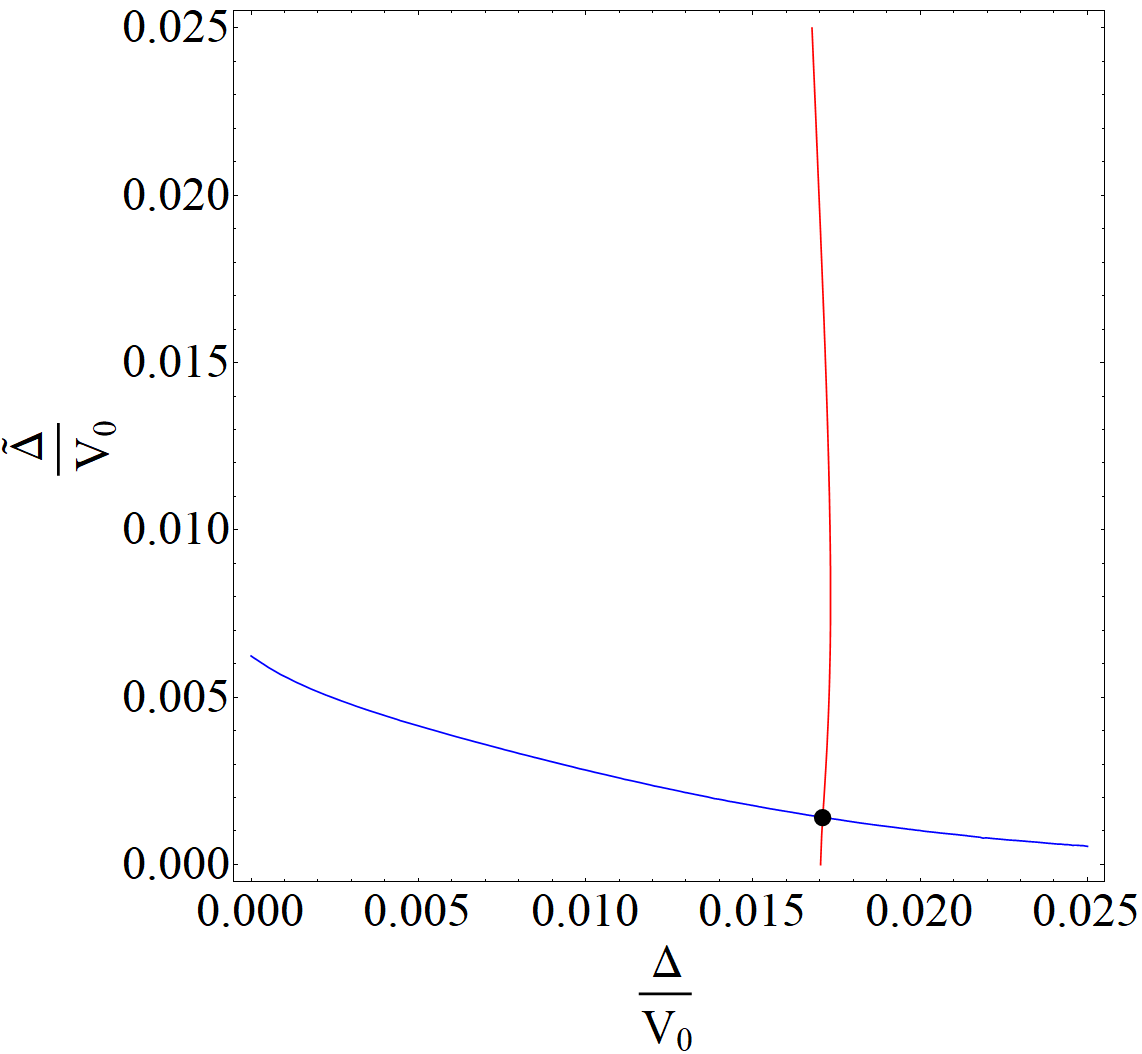}
	\caption{The red line and the blue line represent solutions of Eqs. (\ref{sc1}) and (\ref{sc2}), respectively. The black dot marks the corresponding self-consistent solution.}
	\label{plot1}
\end{figure}

In the LLL, the Eqs. (\ref{sc1}) and (\ref{sc2}) lead to self-consistent solutions with
the following amplitudes, in the kinetic part, $ | {\Delta}_{\bs k}| = \Delta  |\bs k |$ where
$\Delta \approx 0.017086 \; V_0 $, and in the pairing part, $| {\tilde \Delta}_{\bs k}| = {\tilde \Delta} |\bs k |$ where ${\tilde{\Delta}} \approx 0.001414\; V_0 $. The self-consistent solution is numerically obtained using \textit{Mathematica} with error estimation to $ 5 \times 10^{-7} $. Obviously this
represents a weak coupling case in which the kinetic part dominates.
Fig. \ref{plot1} illustrates solutions of Eqs. (\ref{sc1}) and (\ref{sc2}) with the corresponding self-consistent solution. \\[0.2cm]

\begin{figure}
	\centering
	\includegraphics[scale=.18]{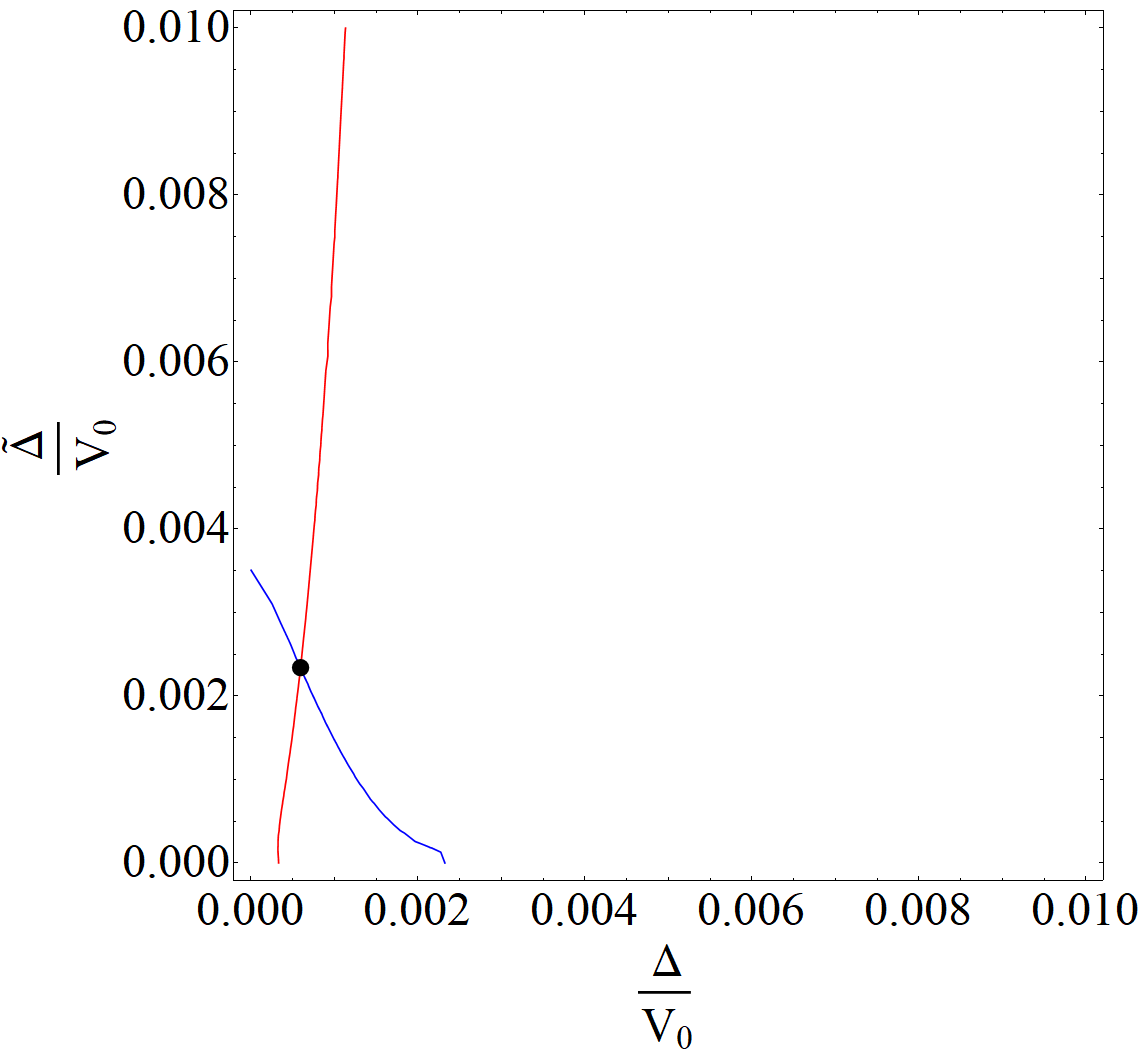}
	\caption{The red line and the blue line illustrate solutions of modified Eqs. (\ref{sc1}) (with $ V \rightarrow \frac{V}{3!}$, etc.) and (\ref{sc2}), respectively. The black dot shows the corresponding self-consistent solution with pairing $| {\tilde \Delta}_{\bs k}| = {\tilde \Delta} |\bs k |$, and $ | {\Delta}_{\bs k}| = \Delta  |{\bs k}|^3 $.}
	\label{plot2}
\end{figure}

In the sLL, we found that Eqs. (\ref{sc1}) and (\ref{sc2}) do not support a coexistence of (non-zero)
kinetic and pairing amplitudes, and thus, if only pairing is present it leads to a gapless (critical) $p$-wave state at this level of approximation. We considered the question of coexistence when a cubic term is
generated in the expansion (\ref{expansion}), in the kinetic part. The resulting equations are slightly
modified Eqs. (\ref{sc1}) and (\ref{sc2}) ($ V \rightarrow \frac{V}{3!}$, etc.), and lead to a solution
 that describes a coexistence of pairing, $| {\tilde \Delta}_{\bs k}| = {\tilde \Delta} |\bs k |$ where ${\tilde{\Delta}} \approx 0.002337\; V_0 $, and  now $ | {\Delta}_{\bs k}| = \Delta  |{\bs k}|^3 $ where
$\Delta \approx 0.000591 \; V_0 \; (l_B = 1)$. The numerically obtained solution using \textit{Mathematica} is shown in Fig. \ref{plot2}. The error is estimated to $ 1 \times 10^{-6}$. Thus, this is a strong coupling, weak pairing case that can be identified
with the usual Pfaffian-state case in which all composite fermions are paired in the same way of a $p$-wave.
\bigskip

\section{Conclusions}


To conclude, we presented a microscopic derivation of the Dirac CF theory and proposed an effective (long-wavelength) noncommutative field-theoretical description. The developed formalism can serve as a base
for a deeper understanding of the strongly correlated physics of the half-filled LL. The microscopic formulation is
used to describe (in a mean-field approach) pairing instabilities of the half-filled LL. The description is
consistent with the known experimental and numerical phenomenology and provides further insights into the
nature of pairing instabilities. 



\appendix

\setcounter{section}{0}

\section{Basic structure of noncommutative gauge field theory}

Noncommutative (NC) spaces are geometric structures for which the notion of a point loses its meaning. They arise naturally in physics, the most famous example being the ``quantum'' phase space of generalized coordinates and momenta, understood as mutually non-commuting (incompatible) observables that satisfy Heisenberg's uncertainty relations $[q^{i},p_{j}]=i\hbar\delta^{i}_{j}$. Another important example of NC space is realized by a single particle moving in the LLL; the guiding center coordinates of such a particle satisfy the following non-trivial commutation relation, $[R_{x},R_{y}]=-il_{B}^{2}$. Finally, in the context of quantum gravity, the idea of NC geometry is applied to the structure of space-time itself. In this case, the degree to which space-time coordinates fail to commute is usually taken to be proportional to the square of the Planck length. For a more comprehensive review see, for example, \cite{Szabo}.   
   
Here we give a short review of the basic structure of a noncommutative gauge field theory based on the Seiberg-Witten (SW) construction, first presented in the context of String Theory \cite{SW_map1}. An NC field theory is a field theory defined on a noncommutative space-time, i.e., a space-time described by mutually non-commuting coordinates, $[x^{\mu},x^{\nu}]\neq 0$. One way to implement this NC structure is to deform the algebra of functions (fields) on ordinary (commutative) space-time, by introducing an NC star-product, the simplest of which is the Moyal-Weyl star-product,
\begin{equation}
f(x)\star g(y)=e^{\frac{i}{2}\theta^{\mu\nu}\frac{\partial}{\partial x^{\mu}}\frac{\partial}{\partial y^{\nu}}}f(x)g(y)\vert_{y\rightarrow x}.
\end{equation} 
The deformation parameters $\theta^{\mu\nu}$ make a constant anti-symmetric matrix (hence the term \emph{constant noncommutativity}). Note that the first term in the expansion (in powers of $\theta$) is just the ordinary commutative product of functions; higher order terms represent NC corrections. When applied to coordinate functions themselves, the above formula gives us 
\begin{equation}
[x^{\mu},x^{\nu}]=x^{\mu}\star x^{\nu}-x^{\nu}\star x^{\mu}=i\theta^{\mu\nu},
\end{equation}  
which is the simplest form of noncommutativity. Other types of noncommutativity are related to different star-products, (see \cite{ACD}), but we will be working only with the Moyal-Weyl star-product. A more comprehensive account on various aspects of NC geometry and NC gauge field theory can be found in \cite{NCbook}.

Let $\{T_{A}\}$ be a set of generators of a gauge group $\mathcal{G}$. Infinitesimal SW gauge variation (NC gauge transformation) of an NC field $\Phi$ transforming in the fundamental representation of the gauge group (e.g. a matter field) is defined by
\begin{equation}
\delta^{SW}_{\Lambda}\Phi=i\Lambda\star\Phi,\label{SW_transf}
\end{equation} 
where $\Lambda$ stands for an NC gauge parameter. This transformation rule is analogous to the familiar one in ordinary, commutative gauge field theory, except for the noncommutative Moyal-Weyl star-product.

There is a notorious problem concerning the closure axiom for SW gauge transformations. Namely, assuming that NC gauge parameter is Lie algebra valued, $\Lambda=\Lambda^{A}T_{A}$, the commutator of two SW transformations is given by
\begin{align}
&[\delta^{SW}_{1}\ds\delta^{SW}_{2}]\Phi=(\Lambda_{1}\star\Lambda_{2}
-\Lambda_{2}\star\Lambda_{1})\star\Phi \label{NC_commutator}\\
&= \frac{1}{2}\left([\Lambda^{A}_{1}\ds\Lambda_{2}^{B}]
\{T_{A},T_{B}\}
+\{\Lambda^{A}_{1}\ds\Lambda_{2}^{B}\}[T_{A},T_{B}]\right)\star\Phi.\nn 
\end{align}
Due to the appearance of anticommutator $\{T_{A},T_{B}\}$, infinitesimal SW transformations do not in general close in the Lie algebra of a gauge group. A way to surmount this difficulty is to assume that NC gauge parameter $\Lambda$ belongs to the universal enveloping algebra (UEA), which is always infinite dimensional \citep{Enveloping}. This, however, leads to an infinite tower of new degrees of freedom (new fields), which is not a preferable property. To see this, consider a SW variation of the covariant derivative
\begin{equation}
\delta^{SW}_{\Lambda}D_{\mu}\Phi=i\Lambda\star D_{\mu}\Phi,
\end{equation}
with $D_{\mu}\Phi=\partial_{\mu}\Phi-i V_{\mu}\star\Phi$ (here we only consider ``left'' gauge field; the case of ``right'' gauge field is treated in a similar manner). This implies the following transformation rule for the NC gauge potential,
\begin{equation}
\delta^{SW}_{\Lambda}V_{\mu}=\partial_{\mu}\Lambda+i[\Lambda\ds V_{\mu}],\label{SW_potential}
\end{equation}
meaning that it is also an UEA-valued object, which leaves us with an infinite number of new fields in the theory (one for each basis element of UEA).

SW map resolves this issue by demanding that NC fields can be expressed in terms of the NC parameter $\theta^{\mu\nu}$, commutative gauge parameter $\lambda=\lambda^{A}T_{A}$, commutative gauge potential $v_{\mu}=v_{\mu}^{A}T_{A}$, and their derivatives,
\begin{align}
\Lambda&=\Lambda(\theta, \lambda,v_{\mu};\partial\lambda_{\mu},
    \partial v_{\mu},\dots), \label{NC_parameter}\\
V_{\mu}&=V_{\mu}(\theta, v_{\mu}, \partial v_{\mu}, \dots),    
\end{align}
where the ellipses stand for higher derivatives. In this way, NC theory is defined by the corresponding commutative one. There are no new degrees of freedom, just new interaction terms in the NC action. 

NC gauge transformations are now induced by the corresponding commutative ones,
\begin{align}
\delta^{SW}_{\lambda}\Lambda&=\Lambda(v_{\mu}+\delta_{\lambda}v_{\mu})-\Lambda(v_{\mu}),\\
\delta^{SW}_{\lambda}V_{\mu}&=V_{\mu}(v_{\mu}+\delta_{\lambda}v_{\mu})-V_{\mu}(v_{\mu}),
\end{align}
with $\delta_{\lambda}v_{\mu}=\partial_{\mu}\lambda+i[\lambda,v_{\mu}]$.

From (\ref{NC_commutator}) and (\ref{NC_parameter}) follows a consistency condition for NC gauge parameter,
\begin{equation}
\Lambda_{1}\star\Lambda_{2}-\Lambda_{2}\star\Lambda_{1}+i(\delta^{SW}_{1}\Lambda_{2}-\delta^{SW}_{2}\Lambda_{1})=i\Lambda_{-i[\lambda_{1},\lambda_{2}]},
\end{equation}
which can be solved perturbatively (this makes sense because the star-product is also defined perturbatively). To this end, we represent the NC gauge parameter as an expansion in powers of the NC parameter $\theta$, with coefficients built out of fields from the commutative theory,    
\begin{equation}
\Lambda=\lambda+\Lambda^{(1)}+\Lambda^{(2)}+\dots.
\end{equation} 
The first term in the expansion (zeroth order in $\theta$) is the commutative gauge parameter $\lambda=\lambda^{A}T_{A}$. 

An NC gauge parameter, up to first order in $\theta^{\alpha\beta}$, is given by
\begin{equation}
\Lambda=\lambda-\frac{1}{2}\theta^{\alpha\beta}v_{\alpha}\partial_{\beta}\lambda+\mathcal{O}(\theta^{2}), \label{SW_lambda}
\end{equation}
where $\theta^{\alpha\beta}$ $(\alpha,\beta=0,1,2)$ is a constant antisymmetric matrix of deformation parameters. In our case, $\theta^{\alpha\beta}=\theta\epsilon^{\alpha\beta}$ where $\theta=-l_{B}^{2}$ and $\epsilon^{\alpha\beta}$ is defined by $\epsilon^{0i}=0$ $(i=1,2)$ and $\epsilon^{12}=-\epsilon^{21}=1$. Note that we work in $2+1$ dimensions and that the definition of $\epsilon^{\alpha\beta}$ implies that noncommutativity is realized only between spatial coordinates; it does not involve the time coordinate. Generalization to an arbitrary number of dimensions is straightforward. 

Using the expansion (\ref{SW_lambda}) and the transformation rule (\ref{SW_transf}) we readily obtain
\begin{equation}
\Phi=\phi-\frac{\theta^{\alpha\beta}}{2}v_{\alpha}\partial_{\beta}\phi+\frac{i\theta^{\alpha\beta}}{4}v_{\alpha}v_{\beta}\phi+\mathcal{O}(\theta^{2}).
\end{equation}
Also, from (\ref{SW_potential}) follows the transformation law for the NC gauge field,
\begin{equation}
V_{\mu}=v_{\mu}-\frac{\theta^{\alpha\beta}}{2}v_{\alpha}\left(\partial_{\beta}v_{\mu}+F_{\beta\mu}\right).
\end{equation} 


In connection to our model, with left (L) and right (R) NC $U(1)$ gauge transformations acting on NC CF fields $c$ and $d$, it is convenient to combine the two CFs into a two-component spinor, 
\begin{equation}
\Psi=\begin{pmatrix}
  c  \\
   d   
\end{pmatrix}.
\end{equation}
Under an NC gauge transformation it changes as 
\begin{equation}
\Psi\rightarrow\Psi'=U_{L}\star\Psi\star U_{R}, 
\label{PhiGaugeTr}
\end{equation}
where $U_{L/R}=e^{i\Lambda_{L/R}}$. The NC gauge parameters $\Lambda_{L/R}$ have a special form given by
\begin{equation}
\Lambda_{L/R}=\begin{pmatrix}
 \Lambda_{L/R}^{c} & 0  \\
   0 & \Lambda_{L/R}^{d}   
\end{pmatrix}.
\end{equation}
Infinitesimally, we get
\begin{equation}
\delta\Psi=i\left(\Lambda_{L}\star\Psi+\Psi\star\Lambda_{R}\right).
\end{equation}
Covariant derivative of $\Psi$ is defined by 
\begin{align}
D_{\mu}\Psi&=\partial_{\mu}\Psi-iV^{L}_{\mu}\star\Psi-i\Psi\star V^{R}_{\mu}\label{CovDerivativePsi}
\end{align}
where we introduced left, $V^{L}_{\mu}$, and right, $V^{R}_{\mu}$, NC gauge potential.  
One can show that $D_{\mu}\Psi$ transforms covariantly,
\begin{equation}
(D_{\mu}\Psi)'= U_{L}\star(D_{\mu}\Psi)\star U_{R}, \label{CovDerPsiGaugeTr}
\end{equation}
provided that NC gauge potentials $V_\mu^L$ and $V^{R}_\mu$ transform in the following way:
\begin{eqnarray}
(V^{L}_{\mu})' &=& U_{L}\star V^{L}_{\mu}\star U_{L}^{\dagger}-i\partial_{\mu}U_{L}\star U_{L}^{\dagger} , \nn\\
(V_{\mu}^{R})' &=& U_{R}^{\dagger}\star V_{\mu}^{R}\star U_{R}-iU_{R}^{\dagger}\star\partial_{\mu}U_{R}.\label{a2GaugeTr}
\end{eqnarray}
The NC action (in Euclidean signature) reads as
\begin{equation}\label{A23}
S_{NC}=\int d\tau d^{2}\boldsymbol{x}\left(\Psi^{\dagger}\star\tau^{\mu}D_{\mu}\Psi+i(V^{L}_{0}+V_{0}^{R})\bar{\rho}_e \right),
\end{equation}
where we have 
\begin{equation}
\tau^{0}=\begin{pmatrix} 1 & 0 \\
                            0 & 1
\end{pmatrix},\hspace{0.2cm}  
\tau^{1}=\begin{pmatrix} 0 & i \\
 i & 0\end{pmatrix}, \hspace{0.2cm}
\tau^{2}=\begin{pmatrix} 0 & 1 \\-1 & 0
\end{pmatrix}.
\end{equation}
In terms of components, the action is 
\begin{align}
&S_{NC} = \int  d\tau  d^{2}\boldsymbol{x}\Big( c^{\dagger} \star D_\tau c + d^{\dagger} \star D_\tau d \\
 + &c^\dagger \star (iD_x +D_y)d + d^\dagger \star (iD_x-D_y)c + i (V^{L}_{0}+V_{0}^{R})\bar{\rho}_e \Big). \nn
\end{align}


The SW map allows us to represent NC fields in terms of ordinary fields from the corresponding commutative theory; these commutative fields will be denoted by a hat symbol. NC fields are organized as perturbation series in powers of the deformation parameter, and they reduce to ordinary fields in the $\theta=0$ limit. 

SW expansions of NC gauge parameters $\Lambda_{L/R}$ (up to first order) are (note the sign difference) given by
\begin{align}
\Lambda_{L} &= \hat{\Lambda}_{L} - \frac{\theta}{2}\epsilon^{\alpha\beta}\hat{V}^{L}_\alpha\partial_{\beta}\hat{\Lambda}_{L}, \label{SWLambdaL} \nn \\
\Lambda_{R} &= \hat{\Lambda}_{R} + \frac{\theta}{2}\epsilon^{\alpha\beta}V_{\alpha}^{R}\partial_{\beta}\hat{\Lambda}_{R}.
\end{align}

Likewise, SW expansions of L and R NC gauge potentials are (again, note the sign difference)
\begin{align}
V_{\mu}^{L} &=\hat{V}^{L}_\mu - \frac{\theta}{2}\epsilon^{\alpha\beta}\hat{V}^{L}_\alpha 
\Big(\partial_{\beta}\hat{V}^{L}_\mu+\hat{F}^{L}_{\beta\mu}\Big),  \label{SWVL}\\
V_{\mu}^{R}&=\hat{V}^{R}_\alpha + \frac{\theta}{2}\epsilon^{\alpha\beta}\hat{V}^{R}_\alpha\Big(\partial_{\beta}\hat{V}^{R}_\mu + \hat{F}^{R}_{\beta\mu}\Big),\label{SWVR}
\end{align}
with (commutative) gauge field strengths
\begin{align}
\hat{F}^{L}_{\alpha\beta}& =\partial_{\alpha}\hat{V}_{\beta}^{L}-\partial_{\beta}\hat{V}_{\alpha}^{L}, \nn\\
\hat{F}^{R}_{\alpha\beta} & = \partial_\alpha \hat{V}^{R}_{\beta} - \partial_\beta \hat{V}^{R}_{\alpha}. 
\end{align}
Finally, the SW expansion of the NC matter field reads as
\begin{equation}
\Psi=\hat{\Psi}-\frac{\theta}{2}\epsilon^{\alpha\beta}\left((\hat{V}^{L}_{\alpha}-\hat{V}^{R}_{\alpha})\partial_{\beta}\hat{\Psi}-i\hat{V}^{L}_{\alpha}\hat{V}^{R}_{\beta}\hat{\Psi}\right).\label{SWPsi}
\end{equation}
\newline
In particular, if we take $V^{L}_{\mu}=A_{\mu}$ and $V^{R}_{\mu}=a_{\mu}-A_{\mu}$ (both $V^{L/R}_{\mu}$ act as scalars), we get covariant derivatives (\ref{Dc}) and (\ref{Dd}), and the previous equation becomes 
\begin{equation}
\Psi=\hat{\Psi}+\frac{\theta}{2}\epsilon^{\alpha\beta}\left((\hat{a}_{\alpha}-2\hat{A}_{\alpha})\partial_{\beta}\hat{\Psi}-i\hat{a}_{\alpha}\hat{A}_{\beta}\hat{\Psi}\right).
\label{SWPsi}
\end{equation}
In general, to compute a first order NC correction for a product of two NC fields, we apply the following formula,
\begin{equation}\label{A31}
(A\star B)^{(1)}=A^{(1)}\hat{B}+\hat{A}B^{(1)}+\frac{i}{2}\theta^{\alpha\beta}\partial_{\alpha}\hat{A}\partial_{\beta}\hat{B},
\end{equation}
where $A^{(1)}$ and $B^{(1)}$ are first order NC corrections of $A$ and $B$, respectively.

Thus, we can readily compute the NC covariant derivative, 
\begin{widetext} 
\begin{align}\label{SWDPsi}
D_{\mu}\Psi=\partial_{\mu}\Psi-iV^{L}_{\mu}\star\Psi-i\Psi\star V_{\mu}^{R}=D_{\mu}\hat{\Psi}+\frac{\theta}{2}\epsilon^{\alpha\beta}\left[(\hat{a}_{\alpha}-2\hat{A}_{\alpha})\partial_{\beta}D_{\mu}\hat{\Psi}-i\hat{a}_{\alpha}\hat{A}_{\beta}D_{\mu}\hat{\Psi}-(\hat{f}_{\alpha\mu}-2\hat{F}_{\alpha\mu})D_{\beta}\hat{\Psi}\right],
\end{align}
\end{widetext}
with commutative field strengths $\hat{f}_{\mu\nu}=\partial_{\mu}\hat{a}_{\nu}-\partial_{\nu}\hat{a}_{\mu}$ and $\hat{F}_{\mu\nu}=\partial_{\mu}\hat{A}_{\nu}-\partial_{\nu}\hat{A}_{\mu}$. Commutative covariant derivative is simply $D_{\mu}\hat{\Psi}=\partial_{\mu}\hat{\Psi}-i\hat{a}_{\mu}\hat{\Psi}$ .

Inserting (\ref{SWVL}), (\ref{SWVR}), (\ref{SWPsi}), and (\ref{SWDPsi}) into the NC action (\ref{A23}) and applying the rule (\ref{A31}), we obtain the commutative action (\ref{Com_action}) and its first-order (in $\theta$) NC correction (\ref{NC_correction}).  

In a similar fashion, if we consider covariant derivatives (\ref{Dc_new}) and (\ref{Dd_new}), i.e., 
\begin{align}
D_{\mu}c&=\partial_{\mu}c-i(a_{\mu}^{(1)}+A_{\mu})\star c-ic\star(a_{\mu}^{(2)}-A_{\mu}), \\
D_{\mu}d&=\partial_{\mu}d-i(a_{\mu}^{(2)}+A_{\mu})\star d-id\star(a_{\mu}^{(1)}-A_{\mu}).
\end{align}
and require $a_{\mu}^{(1)}=a_{\mu}^{(2)}=\tfrac{1}{2}a_{\mu}$ (a full symmetry between physical and unphysical sectors), by the same procedure we obtain action (\ref{NC_Son}) that reproduces the Son's theory at the commutative level.

\section{The structure of NC covariant derivatives} 

Let us assume that the structure of NC covariant derivatives is given by Eqs. (\ref{DDc}) and (\ref{DDd}), i.e.,
\begin{align}
D_{\mu}c&=\partial_{\mu}c-i(a_{\mu}^{(1)}+A_{\mu})\star c-ic\star a_{\mu}^{(2)}, \label{B1}\\
D_{\mu}d&=\partial_{\mu}d-ia_{\mu}^{(2)}\star d-id\star (a_{\mu}^{(1)}-A_{\mu}),\label{B2}
\end{align}
as consistent with the microscopic constraints. 

If we want both (\ref{B1}) and (\ref{B2}) to transform covariantly, i.e.,
\begin{align}
D_{\mu}c&\rightarrow U^{c}_{L}\star D_{\mu}c\star U_{R}^{c}, \\
D_{\mu}d&\rightarrow U^{d}_{L}\star D_{\mu}d\star U_{R}^{d},
\end{align}
we get two separate transformation laws for the gauge field $a_{\mu}^{(2)}$ (in the $c$-case it acts from the right and in the $d$-case from the left):
\begin{align}
a_{\mu}^{(2)}&\rightarrow U^{c\dagger}_{R}\star a_{\mu}^{(2)}\star U_{R}^{c}-iU_{R}^{c\dagger}\star\partial_{\mu}U_{R}^{c},\\
a_{\mu}^{(2)}&\rightarrow U^{d}_{L}\star a_{\mu}^{(2)}\star U_{L}^{d\dagger}-i\partial_{\mu}U_{L}^{d}\star U_{L}^{d\dagger}.
\end{align}
Consistency of the two demands the following constraints:
\begin{align}
U_{L}^{d}&=U_{R}^{c\dagger},\\
U_{R}^{c\dagger}\star\partial_{\mu}U_{R}^{c}&=U_{L}^{d}\star\partial_{\mu}U_{L}^{d\dagger}=\partial_{\mu}U_{L}^{d}\star U_{L}^{d\dagger}.\label{B8}
\end{align}
The second equation implies that $(\partial_{\mu}U_{L}^{d})\star U_{L}^{d\dagger}-U_{L}^{d}\star(\partial_{\mu}U_{L}^{d\dagger})=0$, which is not consistent with $U_{L}^{d}$ being unitary. Thus, we must change the structure of the derivatives. 

Now consider the case when two covariant derivatives have the same structure, 
\begin{align}
D_\mu c &= \partial_\mu c - i  A_\mu \star c  - i c\star b_\mu,\\
D_\mu d &= \partial_\mu d - i  A_\mu \star d  - i d\star b_\mu,
\end{align}
with two gauge fields: the external background field $A_{\mu}$ (which is L gauge field for both fields $c$ and $d$) and $b_{\mu}=a_{\mu}-A_{\mu}$ (which is R gauge field for both fields $c$ and $d$). They transform under gauge transformations in the following way:
\begin{align}
A_{\mu}\rightarrow &U_{L}\star A_{\mu}\star U_{L}^{\dagger}-i(\partial_{\mu}U_{L})\star U^{\dagger}_{L},\\
b_{\mu}\rightarrow &U^{\dagger}_{R}\star b_{\mu}\star U_{R}-iU^{\dagger}_{R}\star(\partial_{\mu}U_{R}).
\end{align}
In the commutative limit, these become simple $U(1)$ gauge transformations,
\begin{align}
A_{\mu}&\rightarrow  A_{\mu}+\partial_{\mu}\varphi_{A},\\
b_{\mu}&\rightarrow  b_{\mu}+\partial_{\mu}\varphi_{b}.
\end{align}
The relevant field $a_{\mu}=b_{\mu}+A_{\mu}$ represents an independent $U(1)$ gauge field with respect to $A_{\mu}$. 

In the setup implied by (\ref{Dc_new}) and (\ref{Dd_new}), i.e., 
\begin{align}
D_{\mu}c&=\partial_{\mu}c-i(a_{\mu}^{(1)}+A_{\mu})\star c-ic\star(a_{\mu}^{(2)}-A_{\mu}), \\
D_{\mu}d&=\partial_{\mu}d-i(a_{\mu}^{(2)}+A_{\mu})\star d-id\star(a_{\mu}^{(1)}-A_{\mu}),
\end{align}
we have to consider four different gauge fields $b^{(1)}_{\mu,\pm}\equiv a_{\mu}^{(1)}\pm A_{\mu}$ and $b^{(2)}_{\mu,\pm}\equiv a_{\mu}^{(2)}\pm A_{\mu}$. Their transformation laws are given by\begin{align}
b_{\mu +}^{(1)}&\longrightarrow U_{L}^{c}\star b_{\mu +}^{(1)}\star U_{L}^{c\dagger}-i\partial_{\mu}U_{L}^{c}\star U_{L}^{c\dagger}, \label{B11}\\b_{\mu -}^{(1)}&\longrightarrow U_{R}^{d\dagger}\star b_{\mu -}^{(1)}\star U_{R}^{d}-iU_{R}^{d\dagger}\star\partial_{\mu}U_{R}^{d},\label{B12} \\b_{\mu +}^{(2)}&\longrightarrow U_{L}^{d}\star b_{\mu +}^{(2)}\star U_{L}^{d\dagger}-i\partial_{\mu}U_{L}^{d}\star U_{L}^{d\dagger}, \label{B13}\\
b_{\mu -}^{(2)}&\longrightarrow U_{R}^{c\dagger}\star b_{\mu -}^{(2)}\star U_{R}^{c}-iU_{R}^{c\dagger}\star\partial_{\mu}U_{R}^{c}.\label{B14}
\end{align}
However, these four fields are not mutually independent, $b^{(1)}_{\mu,+}-b^{(1)}_{\mu,-}=b^{(2)}_{\mu,+}-b^{(2)}_{\mu,-}$, because aside from the external field, the theory can have only two more independent gauge fields, connected with two kinds of constraints. If we require that the difference of gauge fields transforms as a gauge field, we will have $U_{L}^{c}=U_{R}^{d\dagger}$ and $U_{L/R}^{c}=U_{L/R}^{d}$, i.e., possibilities for independent gauge transformations, independent gauge fields, will be reduced to just one, but this would contradict our beginning set-up. That is why we need to take $a_{\mu}^{(1)}=a_{\mu}^{(2)}$, and have two independent gauge transformations and fields $a_{\mu}=2a_{\mu}^{(1)}=2a_{\mu}^{(2)}$ and $A_{\mu}$, in the commutative limit.

\end{document}